\documentclass{article}
\usepackage{arxiv}
\usepackage{cite}
\usepackage{amsmath,amssymb,amsfonts}
\usepackage{algorithmic}
\usepackage{graphicx}
\usepackage{amsfonts}
\usepackage{amssymb}
\usepackage{mathrsfs}
\usepackage{algorithm,algorithmic}
\usepackage{hyperref}
\usepackage{arydshln}
\hypersetup{hidelinks=true}
\usepackage{textcomp}
\usepackage{ntheorem}

\usepackage{graphicx}          % Include this line if your 
\usepackage{cite}
\usepackage{amsmath,amssymb,amsfonts}
\usepackage{algorithmic}
\usepackage{graphicx}
\usepackage{amsfonts}
\usepackage{enumerate}
\usepackage{amssymb}
\usepackage{mathrsfs}
\usepackage{algorithm,algorithmic}
\usepackage{xr-hyper} %必须放在 hyperref 前面
\usepackage{hyperref}
\usepackage{arydshln}
\usepackage{textcomp}

\newtheorem{problem}{Problem}
\newtheorem{assumption}{Assumption}
\newtheorem{definition}{Definition}[section]
\newtheorem{theorem}{Theorem}[section]
\newtheorem{proposition}{Proposition}[section]
\newtheorem{remark}{Remark}[section]
\newtheorem{lemma}{Lemma}[section]
\newtheorem{corollary}{Corollary}[section]

\newtheorem{example}{Example}
\newtheorem*{proof}{Proof}[section]

\title{Robust Output Regulation  of  Uncertain\\  Linear Time-Varying Systems}

\author{Jinmeng Zha 
\And 
Zhen Zhang
\thanks{Corresponding author: Z. Zhang (e-mail: zzhang@tsinghua.edu.cn). The authors are both with the Department of Mechanical Engineering, Tsinghua University, Beijing 100084, China.}}
\begin{document}
\maketitle

\begin{abstract}
Robust output regulation for linear time-varying systems has remained an open problem for decades. 
By augmenting the classical immersion viewpoint, we propose the trajectory-matching system immersion framework.
It reformulates the regulator equation as a forced system and demonstrates that finding an internal model is equivalent to reproducing the non-decaying output trajectories of this forced system  by constructing an unforced one. 
This perspective yields an exact algebraic boundary for finite-dimensional internal models, termed finite linear parameterization.
It further reveals a distinctive obstruction in time-varying systems: even highly structured, finite-dimensional affine parametric uncertainties can generate infinite-dimensional families of non-decaying error-zeroing mappings, thereby precluding exact robust regulation via linear finite-dimensional internal models in general.
Hence, we develop a general finite-dimensional design for approximate robust regulation, which yields a bounded tracking error and avoids explicitly solving the regulator equation.
Additionally, it recovers exact regulation for certain structured uncertainty classes.
Overall, these results clarify the intrinsic limitation of exact finite-dimensional robust regulation for uncertain LTV systems and provide a general and constructive framework for designing LTV internal-model-based regulators.
\end{abstract}

\section{Introduction}
\label{sec-introduction}

Output regulation is one of the central topics in control theory, which studies how to drive the output of a system to track references and/or reject disturbances generated by an exogenous system.
It is completely solved for LTI systems~\cite{davison1976robust,francis1976internal,francis1977linear}, 
and an important outcome is the \textit{internal model principle}, which indicates that asymptotic tracking can only be achieved by containing suitable copies of the exosystem in the controller.
The problem  has been extended to nonlinear systems~\cite{isidori1990output,serrani2002semi,li2025global,jiang2025output}, 
infinite-dimensional systems~\cite{guo2025practical,zhao2025robust}, and
multi-agent systems~\cite{wieland2011internal,lin2025direct,gul2026multi,li2026data}.
Output regulation has found wide applications in electrohydraulic systems~\cite{song2014robust}, electromechanical systems~\cite{cao2025enhanced}, and robotic networks~\cite{liu2025adaptive}.

Output regulation of LTV systems is introduced in~\cite{pinzoni1993output} and investigated with a more comprehensive theory developed by~\cite{zhang2006linear,ichikawa2006output}, which laid the foundation for subsequent research. 
Moreover, inspired by the structure of repetitive control, \cite{sun2009trajectory,zhang2010novel} provide a practical method of constructing an internal-model-based controller without calculating error-zeroing input.
The controller structure is further improved to a parallel one to allow a low-order stabilizer~\cite{zhang2014discrete}; while the low-order stabilizer for the series structure is proposed in~\cite{song2015low}.
In a different vein, \cite{marconi2013internal, cox2016isolating} study the output regulation of hybrid systems for LTI plants with periodic jumps; 
\cite{paunonen2012periodic,paunonen2017robust} extend the problem to periodic linear distributed parameter systems;  and \cite{niu2024adaptive} considers the regulation of non-periodic and discontinuous exogenous signals. 

A major advantage of output regulation is that robustness can be readily achieved for LTI systems~\cite{knobloch1993topics,isidori2003robust}.
Such robustness, however, relies on a crucial property: in the LTI case, the error-zeroing matrix is constant for each fixed uncertainty, and hence the whole uncertain error-zeroing family remains in a finite-dimensional vector space.
For LTV systems, this property does not hold, making robustness substantially more challenging.
The error-zeroing mapping becomes time-varying, so the effect of uncertainty has to be analyzed in a function space rather than only through finite-dimensional algebraic objects such as eigenvalues, frequencies, or constant matrix elements.

Important progress has been made by extending robust designs from LTI cases to LTV ones under additional structural conditions.
For periodic systems, \cite{zhang2006linear,zhang2009tac} construct robust regulators under uniform observability and a Cayley-Hamilton-like property.
For hybrid systems with periodic jumps, \cite{marconi2013internal} develops robust internal-model designs under a separable dependence of the error-zeroing matrix on the uncertainty, and \cite{carnevale2017robust} further provides a semiclassical design with solvability conditions testable on the nominal plant data.
These results establish constructive finite-dimensional designs for important classes of systems.
Another thread of study adopts an infinite-dimensional viewpoint. In particular, \cite{paunonen2017robust} lifts a periodic system to an  infinite-dimensional LTI one, and then  solves the problem by an infinite-dimensional regulator~\cite{paunonen2010internal}.
This lifting-based approach is naturally tailored to periodic systems and provides a robust solution without requiring a finite-dimensional realization. 
At the same time, some finite-dimensional LTV problems may still admit finite-dimensional regulators, motivating a more precise characterization of the boundary between the finite- and infinite-dimensional internal-model realizability.

Recent results on nonlinear output regulation provide a valuable perspective on this issue. In particular, \cite{bin2022robustness} proves the existence of nonlinear regulation problems for which no smooth finite-dimensional regulator can achieve exact robust regulation under arbitrary small continuous perturbations. It suggests that exact robust regulation may be too strong when the uncertainty class is sufficiently rich, motivating us to investigate whether a similar finite-dimensional obstruction appears in uncertain LTV systems. 
This paper demonstrates that even highly structured uncertainties may generate infinite-dimensional families of error-zeroing mappings, which exceed the capability of any linear finite-dimensional internal model.
Furthermore, we identify the high-frequency gain and the zero dynamics as two sources of this ``dimension explosion''. 
This characterization also provides a structural interpretation of the additional assumptions adopted in periodic designs~\cite{zhang2009tac,marconi2013internal}, while being consistent with the use of infinite-dimensional realizations in lifting-based approaches~\cite{paunonen2017robust}. Accordingly, in line with the paradigm shift advocated in nonlinear output regulation~\cite{bin2022robustness,astolfi2022harmonic,bin2024robust}, we argue that  robust regulation of uncertain LTV systems via finite-dimensional  internal models should in general pursue approximate rather than exact regulation.

The main contributions of this paper are listed as follows:
\begin{enumerate}
    \item We systematically solve the LTV regulator equation in a coordinate-free manner, proving that  its solutions are UB and asymptotically unique under standing assumptions. 
    Based on this, we extend the classical system immersion viewpoint from a one-sided signal-generation condition to a \textit{two-system} trajectory-matching characterization, which demonstrates that constructing a robust internal model is equivalent to reproducing the output trajectory of a given forced system via an unforced system.
    \item With the proposed trajectory-matching system immersion, we reveal the fundamental dynamic influence of parametric uncertainties. 
    We derive the exact algebraic \textit{boundary} for finite-dimensional internal models, termed finite linear parameterization. Furthermore, we prove this boundary is generally violated for uncertain LTV systems. In particular, uncertainties may excite infinite-dimensional families of functions, making exact robust regulation via finite-dimensional internal-model-based regulators structurally \textit{unattainable} in general.
    \item The inherent barriers motivate us to pursue approximate regulation instead of exact regulation. 
    We develop a truncation-based finite-dimensional approximate regulator, which avoids explicitly solving the regulator equation. It achieves exact regulation for some specified uncertainties, and ensures approximate regulation with bounded tracking errors for general uncertainties.
\end{enumerate}

\textit{Notations}: The transition matrix of $A(t)$ is given by $\Phi_A(t,t_0)$. As exceptions, the transition matrices of $A_\eta(t)$ and $A_{\rm cl}(t) $ are given by $\Phi_\eta(t,t_0)$ and $\Phi_{\rm cl}(t,t_0)$ respectively for simplicity.
The space of smooth functions from $X$ to $Y$ is denoted $C^\infty (X,Y)$.
$I_N$ is the identity matrix of dimension $N$, and $\otimes$ denotes the Kronecker product.
The subscript ``$0$'' represents the nominal value calculated using nominal parameters.
For brevity, time and parameter dependencies namely $t$ and $\mu$ are omitted where the context is unambiguous.
$\phi_0>1$ denotes a generic constant that may vary from line to line but
depends only on the fixed system data and the compact uncertainty set, which means $\phi_0$ is independent of \(t\), \(\mu\), and the design parameters like gains or approximation orders. 

\textit{Abbreviations}: 
LTV means linear time-varying;
LTI means linear time-invariant;
UAS means uniformly asymptotically stable with respect to time; 
IM means internal model; 
RE means regulator equation;
UB means uniformly bounded with respect to time; for solutions initialized at $t_0$, UB is understood on $t\ge t_0$.

\section{Problem Formulation}
Consider the LTV system
\begin{equation}
\begin{aligned}
\dot w&= S(t) w\\
\dot x&= A(t,\mu) x +B(t,\mu)u + P(t,\mu)w\\
e &= C(t,\mu)x+Q(t,\mu)w,\\
\end{aligned}
\label{sys}
\end{equation}
with exosystem state $w \in \mathbb{R}^{\rho}$, plant state $x \in \mathbb{R}^{n}$, control input $u \in \mathbb{R}$, and regulated error $e \in \mathbb{R}$.  
The parameter vector $\mu$ takes values on a given known compact set $\mathcal{P}\subseteq \mathbb{R}^N$, and all the matrices take the nominal values when $\mu = \mathbf{0}$.
We assume all the known matrices are smooth functions, and their derivatives of necessary orders including themselves are UB.
Let $\mathcal U$ be an open neighborhood of $\mathcal P$. All parameter-dependent mappings are understood to be defined and smooth with respect to $\mu$ on $\mathcal U$, while all 
identities and uniform estimates are required only for $\mu\in\mathcal P$.
Consistent with standard formulations~\cite[Sec. 1.6]{knobloch1993topics} and~\cite{zhang2006linear}, we assume the exosystem dynamics $S(\cdot)$ are precisely known.

The problem of robust time-varying output regulation is defined as follows:
\begin{problem}[Robust Regulation]\label{prob-1}
For the system~\eqref{sys}, find an error-feedback regulator in the form of
\begin{equation}
\begin{aligned}
\dot \xi &= F(t) \xi+ G(t)e\\
u &= H(t) \xi, \\
\end{aligned}
\label{controller}
\end{equation}
with state $\xi \in \mathbb{R}^\nu$, such that 
\begin{enumerate}[(i)]
    \item the unforced closed-loop system ${\rm col}(x,\xi)$, namely
\begin{equation} \nonumber
  \hspace{-2em}
  A_{\rm cl}(t,\mu) = \begin{bmatrix}
      A(t,\mu) & B(t,\mu)H(t) \\
      G(t)C(t,\mu) & F(t) 
    \end{bmatrix}
\end{equation}
is UAS, and
\item the trajectories of the closed-loop system originating from any initial state $(x_0,\,\xi_0,\,w_0) \in \mathbb{R}^{n+\nu+\rho}$ are bounded and satisfy 
\begin{equation}
  \lim_{t\to \infty} e(t) =0, \label{eq-e-rr}
\end{equation}
\end{enumerate}
for all $\mu \in \mathcal{P}=\{||\mu||_\infty \le \phi_\mu\}\subseteq \mathbb{R}^N$, with the given constant $\phi_\mu>0$.
\end{problem}
When Problem~\ref{prob-1} is unattainable, a more practical version is pursued, similar to the nonlinear case~\cite{bin2022robustness}.
\begin{problem}[Approximate Regulation]
  Approximate regulation refers to the case similar to Problem~\ref{prob-1} while replacing the target~\eqref{eq-e-rr}  with 
  \begin{equation}
    \limsup_{t\to \infty}|e(t)|\le \phi_e,  \nonumber
  \end{equation}
  where $\phi_e > 0$ is a constant.
\end{problem}

The exosystem state is assumed to be bounded and non-decaying, namely
\begin{assumption}\label{ass-s}
The exosystem is \textit{marginally stable}, meaning there exists a constant $\tilde \phi_S>0$, such that for any $t,\tau \in \mathbb{R}$, $||\Phi_S(t,\tau)||\leq \tilde \phi_S$.
\end{assumption}
To render the following analysis, we need the system to possess uniform relative degree:
\begin{assumption} \label{ass-r}
The plant model in~\eqref{sys} has \textit{uniform relative degree} $r$~\cite[Def. 5]{de1998zeros}, 
meaning for any $t\in \mathbb{R},\mu \in \mathcal{P}, i=0,\cdots,r-2$, $[L_A^i C(t,\mu)] B(t,\mu)\equiv 0$ (see Supplementary~S1), and there exists a constant $\phi_b>0$ such that the high-frequency gain
\begin{equation}
    b (t,\mu) = [L_A^{r-1} C(t,\mu)]B(t,\mu) \nonumber
\end{equation}
satisfies $|b(t,\mu)|>\phi_b$.
\end{assumption}
Unless specified otherwise, the derivations in this paper are coordinate-free.
Sometimes we adopt the Byrnes-Isidori form for simplicity, denoted by the subscript ``BI'', namely
\begin{equation} \nonumber
  \begin{aligned}
  A_{\rm BI}(t,\mu) &= 
  \left[
\begin{array}{c:c:c}
 \mathbf{0}_{r-1,1} &I_{r-1}& \mathbf{0}_{r-1,n-r}\\
 \hdashline
\multicolumn{3}{c}{\alpha^T(t,\mu)} \\
 \hdashline
  \beta(t,\mu)&\mathbf{0}_{n-r,r-1}&A_\eta(t,\mu)
\end{array}
\right] \\
B_{\rm BI}(t,\mu) &=  
\left[
\begin{array}{c}
 \mathbf{0}_{r-1,1}\\
 \hdashline
b(t,\mu) \\
 \hdashline
  \mathbf{0}_{n-r,1}
\end{array} \right] \\
C_{\rm BI}(t) &=  \begin{bmatrix}
  1 & \mathbf{0}_{r-1,1}^T & \mathbf{0}^T_{n-r,1}
 \end{bmatrix} ,
  \end{aligned}
\end{equation}
which specifies the zero dynamics as $A_\eta(\cdot,\cdot)$ and the high-frequency gain as $b(\cdot,\cdot)$.
We also assume 
\begin{assumption}\label{ass-m}
  The system~\eqref{sys} is topologically equivalent (See Supplementary~S1) to Byrnes-Isidori form, and  the plant model  is \textit{minimum phase}, namely, there exists a parameterized family of symmetric matrix-valued functions $P_{\eta}(t,\mu)$ and constants $a_1,a_2,a_3>0$ such that for all $(t,\mu)\in\mathbb R\times\mathcal P$, $a_1 I \leq P_{\eta}(t,\mu) \leq a_2 I$ and $\dot{P}_{\eta}(t,\mu) + P_{\eta} A_\eta(t,\mu) + A_\eta^T(t,\mu) P_{\eta}(t,\mu) \leq -a_3I$.
\end{assumption}
This implies $||\Phi_{\eta} (t,\tau,\mu)||\leq \tilde \phi_\eta e^{-\hat \phi_\eta (t-\tau)}$, for any $ t\geq \tau$, where the constants $\tilde \phi_\eta=\sqrt{a_2/a_1}$ and $\hat \phi_\eta=a_3/2a_2$.
While Assumption~\ref{ass-r} ensures the algebraic existence of Byrnes-Isidori form~\cite[Thm. B.7]{berger2015zero}, Assumption~\ref{ass-m} is needed to guarantee the transformation is Lyapunov,
unless the system  is periodic~\cite[Thm. 8]{de1998zeros}. 

Define
\begin{equation}
  \mathscr{O}_A(t,\mu) = 
  \left [
  \begin{matrix}
  L_A ^0 C(t,\mu)\\
  L_A^ 1 C(t,\mu)\\
  \vdots \\
    L_A^{r-1} C(t,\mu)
  \end{matrix}
  \right ],
  \nonumber
\end{equation}
and
\begin{equation}
  \bar {\mathscr{O}}_A (t,\mu) = L_A^{r} C(t,\mu) .\label{eq-oa} \nonumber
\end{equation}
The definitions of $\mathscr{O}_S(\cdot,\cdot)$ and $\bar {\mathscr{O}}_{S}(\cdot,\cdot)$ are similar by directly replacing $\{A(\cdot,\cdot),C(\cdot,\cdot)\}$ with $\{S(\cdot),Q(\cdot,\cdot)\}$ while keeping $r$ invariant. Then define 
  $\mathscr{P}^T  = 
  \begin{bmatrix}
  \mathscr{P}_1^T&
  \mathscr{P}_2^T&
  \cdots &
  \mathscr{P}_r  ^T
  \end{bmatrix}$,
  $  \bar {\mathscr{P}}  = \mathscr{P}_{r+1}$,
where $\mathscr{P}_1=0$, and
\begin{equation}
  \begin{aligned}
      \mathscr{P}_i =& L_S ^0(L_A^{i-2} C ) P +L_S^1 ( (L_A^{i-3} C ) P)+\cdots 
      + L_S ^{i-2} ( (L_A^{0} C ) P  ), i = 2,3,\cdots,r+1. \nonumber
  \end{aligned}
\end{equation}
Define 
\begin{equation}
    \begin{aligned}
        M   &= A - b ^{-1}B \bar {\mathscr{O}}_A ,\\
        N  &= -  b ^{-1} B  (\bar {\mathscr{O}}_{S} + \bar {\mathscr{P}} )+P ,\\
        \bar N  &= -  b ^{-1}  (\bar {\mathscr{O}}_{S} + \bar {\mathscr{P}} ).        
    \end{aligned}\label{abbrs1}  \nonumber
\end{equation}
Denote the upper $r$ rows and the remaining lower rows of a matrix by the subscripts ``${\rm u}$'' and ``${\rm l}$'', respectively.
Additionally, some background is provided in Supplementary~S1.

\section{Trajectory-Matching System Immersion}\label{sec-imi}

The results in  \cite[Prop. 3.2]{zhang2006linear} and \cite[Thm. 2.1]{ichikawa2006output} indicate the conditions for solving Problem~\ref{prob-1}.
\begin{proposition}\label{pr-re}
With Assumption~\ref{ass-s}, suppose the regulator~\eqref{controller} has stabilized the closed-loop system. 
Then, the following statements hold:

\begin{enumerate}[(i)]
  \item Necessary and Sufficient Condition:
The regulator solves Problem~\ref{prob-1} if and only if there exist UB mappings $  \Pi\in C^{\infty}(\mathbb{R}\times \mathcal{U}, \mathbb{R}^{n\times\rho})$, $   \Sigma\in C^{\infty}(\mathbb{R}\times \mathcal{U} , \mathbb{R}^{\nu\times\rho})$ and $R\in C^{\infty}(\mathbb{R}\times \mathcal{U} ,\mathbb{R}^{1 \times\rho})$, such that for all $(t,\mu)\in\mathbb R\times\mathcal P$, (omitting arguments)
\begin{equation}
\begin{aligned}
\dot {  \Pi}+ {  \Pi}S &= A \Pi+BR+P \\
\mathbf{0}&=\lim_{t\to\infty} (C{ \Pi}+Q ),
\end{aligned} \label{REiff1}
\end{equation}
and 
\begin{equation}
    \begin{aligned}
\dot {  \Sigma}+ {  \Sigma} S &= F  \Sigma + G(C  \Pi +Q )\\
R &= H  \Sigma .
\end{aligned} \label{REiff2}
\end{equation}

\item Sufficient Condition:
The regulator solves Problem~\ref{prob-1} if there exist UB mappings $\bar  \Pi\in C^{\infty}(\mathbb{R}\times \mathcal{U}, \mathbb{R}^{n\times\rho})$, $  \bar \Sigma\in C^{\infty}(\mathbb{R}\times \mathcal{U} , \mathbb{R}^{\nu\times\rho})$ and $\bar R\in C^{\infty}(\mathbb{R}\times \mathcal{U} ,\mathbb{R}^{1 \times\rho})$ satisfying the strengthened equations
\begin{align}
\dot {\bar  \Pi}+ { \bar \Pi}S &= A\bar \Pi+B \bar R+P\label{RE1} \\
\mathbf{0}&= C{\bar \Pi}+Q,\label{RE2}
\end{align}
and
\begin{align}
\dot { \bar \Sigma} + { \bar \Sigma} S &= F \bar \Sigma\label{RE3}\\
\bar R &= H \bar \Sigma\label{RE4},
\end{align}
for all $(t,\mu)\in\mathbb R\times\mathcal P$.
If the system is $T$-periodic, the conditions that \eqref{RE1}-\eqref{RE4} admit a $T$-periodic solution become necessary and sufficient. 
\end{enumerate}
\end{proposition}

Proposition~\ref{pr-re} transforms the problem of robust regulation into the solvability of the constrained Sylvester differential equations.
In detail, condition (i) and (ii) each  include four equations. 
The first two equations, namely~\eqref{REiff1}, or~\eqref{RE1}-\eqref{RE2} when dropping the ``limit'', are termed the \textit{regulator equations}, which completely depend on the system~\eqref{sys}.
Hence, their solvability is a necessary condition for that of Problem~\ref{prob-1}, which is studied in Section~\ref{sec-re} with its solution demonstrated.
The remaining equations~\eqref{REiff2} or~\eqref{RE3}-\eqref{RE4} depend on the regulator, which reveals that the essence of robust regulation is to design regulator parameters so that~\eqref{REiff2} or~\eqref{RE3}-\eqref{RE4} are always solvable. This procedure is reformulated as the proposed trajectory-matching system immersion in Section~\ref{sec-tsi}.

\subsection{Solution to the Time-Varying Regulator Equation}\label{sec-re}
\begin{theorem}\label{thm-re}
With Assumptions~\ref{ass-s}-\ref{ass-m}, the UB solutions 
%$\{\Pi(t,\mu),R(t,\mu)\}$ 
to the RE~\eqref{REiff1} always exist, one of them is 
\begin{align}
   \bar \Pi(t,\mu) =&  \Phi_M(t,t_0,\mu)\bar\Pi(t_0,\mu)\Phi_S(t_0,t)    
    + \int_{t_0}^{t} \Phi_M(t,\tau,\mu) N(\tau,\mu)\Phi_S(\tau,t)  \,\mathrm{d}\tau \label{eq-pi}\\
 \bar R(t,\mu) =& - b^{-1} (t,\mu) \bar {\mathscr{O}}_A (t,\mu)\bar \Pi(t,\mu)+\bar N(t,\mu) ,\label{eq-R} 
\end{align}
with the initial value being any solution to
\begin{equation}
     \mathscr{O}_{A}(t_0,\mu)\bar \Pi (t_0,\mu)+ \mathscr{O}_{S}(t_0,\mu)+\mathscr{P}(t_0,\mu) =0. \label{eq-iv}
\end{equation}
Moreover, all the solutions $\{\Pi(\cdot,\cdot),R(\cdot,\cdot)\}$ converge to~\eqref{eq-pi}-\eqref{eq-R} when $t\to \infty$, if   the first $k$ derivatives of $R(\cdot,\cdot)$ are UB with $k=\max\{1,r-2\}$.
Additionally, one of the solutions in Byrnes-Isidori form is
  \begin{align}
  \bar  \Pi_{\rm BI,u}(t,\mu) =& - (\mathscr{O}_S(t,\mu) + \mathscr{P}(t,\mu)) \label{eq-pi-BIu}\\
\bar \Pi_{\rm BI,l}(t,\mu) =&\int_{t_0}^{t} \Phi_{\eta}(t,\tau,\mu) ( P_{\rm BI, l}(\tau,\mu) - \beta(\tau,\mu) Q(\tau,\mu))\Phi_S(\tau,t)  \,\mathrm{d}\tau .\label{eq-pi-BI}
  \end{align}
\end{theorem}

\begin{proof}
See~\ref{pf-thm-re} for a proof.
\end{proof}

\begin{remark}\label{rm-re}
The solution~\eqref{eq-pi}-\eqref{eq-R} is also the solution to the strengthened RE~\eqref{RE1}-\eqref{RE2}.
Though its solution space narrows compared to the original RE~\eqref{REiff1}, it is still reasonable to remove the ``limit'' when solving the RE,
as the asymptotic constraint in~\eqref{REiff1} is analytically prohibitive for solvability.
Moreover, the solution to~\eqref{RE1}-\eqref{RE2} 
is representative enough, because all the solutions to the original RE converge to it.
This coincides with the intuition that we only care about the performance when $t$ is large enough.
\end{remark}

\begin{remark}
  To the best of our knowledge, \eqref{eq-pi}-\eqref{eq-R} provide a coordinate-free representation of the UB solution to the LTV RE for the first time. When expressed in Byrnes-Isidori coordinates, it recovers the solution structure reported in~\cite[Thm. 2]{shim2006output}.
  Assumption~\ref{ass-m} ensures UB of the solution, and other assumptions, such as an exponential dichotomy~\cite{shim2010note}, can also be used. 
  The condition on the first $k$ derivatives of $R(\cdot,\cdot)$ is to ensure asymptotic uniqueness, which can be  replaced by requiring $\dot R(\cdot,\cdot)$ to be UB, with $A(\cdot,\cdot)$ UAS (See Lemma~\ref{lm-au}).
  Moreover, a whole-line UB representative is obtained from~\eqref{eq-pi-BIu}-\eqref{eq-pi-BI} by setting $t_0=-\infty$.
  In the periodic case, it is the unique $T$-periodic solution, so the derivative condition can be omitted.
  Indeed, the lower part namely $\bar \Pi_{\rm BI,l}(\cdot,\cdot)$  is the unique solution to~\eqref{eq-se-sim} by~\cite[Lem. A.1]{zhang2006linear}.
  More discussion on uniqueness is shown in Remark~\ref{rm-au}.
\end{remark}

\subsection{System Immersion via Trajectory Matching}\label{sec-tsi}

After solving \eqref{REiff1} in Theorem~\ref{thm-re}, this section turns to  the solvability of~\eqref{REiff2}, with particular focus on the strengthened equations~\eqref{RE3}-\eqref{RE4}.
A straightforward way in the LTI case is to follow the separation principle~\cite[Sec. 1.4]{knobloch1993topics}.
However, for  LTV robust regulation, another way called \textit{system immersion} (See Supplementary~S1) seems more promising, which is adopted in~\cite[Sec. 1.4]{isidori2003robust} for the LTI case  and~\cite{zhang2009tac,marconi2013internal} for the periodic one.
In the system-immersion approach, the uncertain family of
signal-generating pairs $\{S(\cdot),R(\cdot,\mu)\}_{\mu\in\mathcal P}$ is reproduced by a parameter-independent unforced system~$\{\Xi(\cdot),\Gamma(\cdot)\}$, which is referred to as an \textit{internal model}. 

\begin{definition}[Internal Model]\label{def-im}
  Given $S\in C^\infty(\mathbb{R}, \mathbb{R}^{\rho\times \rho})$ and $\Xi\in C^\infty(\mathbb{R}, \mathbb{R}^{\bar \nu\times \bar \nu})$ both marginally stable, with UB mappings $R\in C^{\infty}(\mathbb{R}\times \mathcal{U} ,\mathbb{R}^{1 \times\rho})$ and $\Gamma\in C^\infty (\mathbb{R} ,\mathbb{R}^{1\times \bar \nu})$,
  the pair $\{\Xi(\cdot),\Gamma(\cdot)\}$ is an (exact) \textit{internal model} for $\{S(\cdot),R(\cdot,\mu)\}$, if for all $\mu \in \mathcal{P}$,
 there exists a UB mapping  $\hat \Sigma(t,\mu)\in C^\infty(\mathbb{R}\times\mathcal U, \mathbb{R}^{\bar\nu\times\rho})$  such that 
    \begin{align}
\dot { \hat \Sigma}(t,\mu) + { \hat \Sigma}(t,\mu) S(t) = \Xi(t) \hat \Sigma(t,\mu),\label{eq-imdef1}
\end{align}
and $\Delta_R(t,\mu) \equiv 0$, where 
\begin{equation}
  \Delta_R(t,\mu) = R(t,\mu) - \Gamma(t) \hat \Sigma(t,\mu).\label{eq-imdef2}
\end{equation}
$\{\Xi(\cdot),\Gamma(\cdot)\}$ is called an \textit{asymptotic internal model}, if for all $\mu \in \mathcal{P}$,  there exists a UB mapping  $\hat \Sigma(t,\mu)\in C^{\infty}(\mathbb{R}\times\mathcal{U}, \mathbb{R}^{\bar \nu \times \rho})$ satisfying~\eqref{eq-imdef1}, and
\begin{equation}
  \lim_{t\to \infty} \Delta_R(t,\mu)=\mathbf{0}.\label{eq-as-im}
\end{equation}
It is called an \textit{approximate internal model}, if $\Delta_R(t,\mu)$ is UB.
In particular, exact and asymptotic IMs are special cases of approximate IMs. 
Moreover, when the system is $T$-periodic, $\{\Xi(\cdot),\Gamma(\cdot)\}$ is also required to be $T$-periodic, and then an asymptotic IM is indeed an exact IM.

Two IMs are called equivalent, if they both are IMs for $\{S(\cdot),R(\cdot,\mu)\}$.
An IM is called \textit{minimum} if 
  \begin{enumerate}[(i)]
    \item $\{\Xi(\cdot),\Gamma(\cdot)\}$ is completely observable.
    \item  for any $t_0\in\mathbb{R}$ and nonzero $\lambda \in \mathbb{R}^{\bar \nu}$,  there exists $\mu\in \mathcal{P}$, such that $ \lambda^T \hat \Sigma(t_0,\mu) \neq 0$.
  \end{enumerate}
An IM is called \textit{regular} if $\{\Xi(\cdot),\Gamma(\cdot)\}$ is uniformly completely observable.
\end{definition}

Throughout this paper, an IM is understood to be admissible in the sense that it and its derivatives of the necessary orders are UB.
Definition~\ref{def-im} explains why the solvability of~\eqref{RE3}-\eqref{RE4} can be studied through IMs. Indeed,~\eqref{eq-imdef1}-\eqref{eq-imdef2} with $\Delta_R(\cdot,\cdot)\equiv0$ have the same structure as~\eqref{RE3}-\eqref{RE4}, with the IM triple $\{\Xi(t),\Gamma(t),\hat\Sigma(t,\mu)\}$ replacing the regulator triple $\{F(t),H(t),\bar\Sigma(t,\mu)\}$.
A regular IM will later be embedded into a stabilizing regulator $\{F(\cdot),G(\cdot),H(\cdot)\}$ in~\ref{sec-st}, where an IM is said to be \textit{embedded} in a regulator if it is immersed into the regulator's zero-input pair $\{F(\cdot),H(\cdot)\}$ through a UB immersion mapping (See Supplementary~S1).
Hence, the final regulator still admits a mapping~$\bar \Sigma(\cdot,\cdot)$ satisfying~\eqref{RE3} with $H(\cdot)\bar \Sigma(\cdot,\cdot)$ reproducing the output of the embedded IM.
Detailed discussion on IMs is provided in~\ref{app-im}, including the interpretations of minimum and regular IMs in Remark~\ref{rm:cim}, the reduction to a minimum IM, and the construction of equivalent IMs in Remark~\ref{rm-eqim}. 
Approximate IMs are studied in Section~\ref{sec-su}.
Therefore, we focus on designing IMs hereafter.

By reformulating~\eqref{RE1}-\eqref{RE4}, system immersion is augmented as Theorem~\ref{thm-tsi}, which reveals the equivalence between designing robust IMs and matching the trajectories of two systems. 
Denote $\bar \Psi(t,t_0,\mu) = \bar \Pi(t,\mu) \Phi_S(t,t_0)$,  
  $\bar \Lambda(t,t_0,\mu) =\bar \Sigma(t,\mu) \Phi_S(t,t_0)$, and
  $\bar\Omega(t,t_0,\mu) =\bar R(t,\mu)\Phi_S(t,t_0)$.

\begin{theorem}[Trajectory-Matching System Immersion]\label{thm-tsi}
 With Assumptions~\ref{ass-s}-\ref{ass-m}, the equations~\eqref{RE1}-\eqref{RE4} are solvable, if and only if trajectory-matching system immersion is feasible. This means one can construct an unforced system $\{F(\cdot),H(\cdot)\}$ given by 
\begin{equation}
\begin{aligned}
  \dot {\bar\Lambda} (t,t_0,\mu)  &= F(t)\bar \Lambda (t,t_0,\mu)\\
   \bar \Omega(t,t_0,\mu) &= H(t) \bar\Lambda (t,t_0,\mu),
    \end{aligned}\label{eq-isi}
\end{equation}
 such that it can reproduce all the  output trajectories of the given forced system in the presence of uncertainty, namely
  \begin{equation}
    \begin{aligned}
         \dot {\bar\Psi }(t,t_0,\mu)  &= M(t,\mu)\bar \Psi(t,t_0,\mu) +N(t,\mu)\Phi_S(t,t_0) \\
         \bar\Omega(t,t_0,\mu) &= - b^{-1} (t,\mu)\bar {\mathscr{O}}_A (t,\mu) \bar\Psi(t,t_0,\mu)
         +\bar N(t,\mu)\Phi_S(t,t_0) ,
    \end{aligned}\label{eq-isi2}
\end{equation}
with constrained initial value satisfying
\begin{equation}
  \mathscr{O}_{A}(t_0,\mu) \bar\Psi (t_0,t_0,\mu)+ \mathscr{O}_{S}(t_0,\mu)+\mathscr{P}(t_0,\mu) =0 .\nonumber
\end{equation}
\end{theorem}

\begin{proof}
Right-multiply \eqref{RE3}-\eqref{RE4} and \eqref{eq-pi}-\eqref{eq-iv} by $\Phi_S(t,t_0)$ to yield the results.
\end{proof}

\begin{remark}\label{rm-tsi}
  Trajectory-matching system immersion extends the classical immersion viewpoint by explicitly characterizing both sides of the immersion. The unforced system~\eqref{eq-isi} corresponds to the classical IM generator, whereas the forced system~\eqref{eq-isi2} is derived from the solution of the RE and describes the trajectory family to be immersed.
  This two-system reformulation provides a deeper insight into the robust regulation problem and serves as a guideline for the subsequent dimensionality analysis.
  Since every UB solution $R(\cdot,\cdot)$ of the original RE~\eqref{REiff1} converges to $\bar R(\cdot,\cdot)$ by Theorem~\ref{thm-re}, it is reasonable to omit the bar when discussing the non-decaying error-zeroing family hereafter.

  \begin{enumerate}[(i)]
    \item
The forced system~\eqref{eq-isi2} makes explicit the dynamical mechanism that generates $R(\cdot,\cdot)$, allowing its structural properties to be investigated  before explicitly solving the RE. In particular, the non-decaying error-zeroing family to be embedded in the IM is not determined by the exosystem alone; it is also shaped by the plant dynamics, especially the high-frequency gain $b(\cdot,\cdot)$ and the zero dynamics $A_\eta(\cdot,\cdot)$.
This viewpoint complements some existing immersion formulations that impose prescribed structures on $R(\cdot,\cdot)$. Indeed, Section~\ref{sec:gid} shows that such structural assumptions can be highly restrictive, since the non-decaying component of the family $\{R(\cdot,\mu):\mu\in\mathcal P\}$ may span an infinite-dimensional function space in general.

  \item The unforced system~\eqref{eq-isi} represents the IM generator embedded in the regulator that generates the error-zeroing input $R(t,\mu)w$. In the design procedure, however, we first construct an abstract IM
$\{\Xi(\cdot),\Gamma(\cdot)\}$ that is later embedded into $\{F(\cdot),H(\cdot)\}$. 
$\{\Xi(\cdot),\Gamma(\cdot)\}$ can be expressed as 
\begin{equation}
  \begin{aligned}
  \dot {\hat \Lambda} (t,t_0,\mu)  &= \Xi(t)\hat \Lambda (t,t_0,\mu)\\
   \bar \Omega(t,t_0,\mu) &= \Gamma(t) \hat\Lambda (t,t_0,\mu),
    \end{aligned}\label{eq-isi-im}
\end{equation}
with $\hat \Lambda(t,t_0,\mu) = \hat \Sigma(t,\mu) \Phi_S(t,t_0)$.
  This also characterizes the finite-dimensional signal-generating capability  of a linear IM, so if Problem~\ref{prob-1} is solvable through an IM,
  $R(\cdot,\cdot)$ has to satisfy the conditions in Section~\ref{sec:flp}.
  \end{enumerate}
\end{remark}

\section{Dimensionality of Internal Models for Robustness}\label{sec:dimr}

Following Remark~\ref{rm-tsi},
Section~\ref{sec:flp} demonstrates the limit of robustness that a linear finite-dimensional IM can achieve.
Then Section~\ref{sec:gid} indicates that  in general, the uncertainty-induced error-zeroing family is beyond its capability, making it impossible to solve Problem~\ref{prob-1} by an IM-based  regulator.
To isolate the dynamic effects of uncertainties, whenever a proposition specifies a perturbed matrix, all other system matrices are assumed independent of $\mu$.

\subsection{Finite Linear Parameterization: The Boundary for Finite-Dimensional Internal Models}\label{sec:flp}

Since all UB solutions of the RE are asymptotically equivalent by Theorem~\ref{thm-re}, an LTV IM only needs to reproduce the asymptotically error-zeroing input rather than an exact one to solve Problem~\ref{prob-1}.
This is different from the LTI and periodic settings, where there is no transient ambiguity of the solutions to the RE.  
Accordingly, the relevant signal generator for LTV systems is an asymptotic IM rather than necessarily an exact one.
However, an asymptotic IM is available only in restricted cases, as shown in Theorem~\ref{thm-flp}. 

\begin{theorem}[Finite Linear Parameterization]\label{thm-flp}
With Assumptions~\ref{ass-s}-\ref{ass-m}, suppose the first $k$ derivatives of the error-zeroing mapping $R(\cdot,\cdot)$ are UB with $k=\max\{1,r-2\}$.
Then there exists a finite-dimensional asymptotic internal model $\{\Xi(\cdot), \Gamma(\cdot)\}$ for $\{S(\cdot),R(\cdot,\mu)\}$, if and only if $R(\cdot,\cdot)$
admits the \textit{finite linear parameterization}, i.e., it can be explicitly parameterized as
\begin{equation}
    R(t,\mu) = R_0(t) + \sum_{k=1}^{N_R} \theta_{Rk}(\mu) R_k(t) + \Delta_{\rm d}(t,\mu), \label{eq-flp}
\end{equation}
where $R_0\in C^{\infty}(\mathbb{R},\mathbb{R}^{1 \times \rho})$ is the UB nominal error-zeroing matrix, $R_k\in C^{\infty}(\mathbb{R},\mathbb{R}^{1 \times \rho})$ are UB and non-decaying functions, $\theta_{Rk}\in C^\infty(\mathcal U,\mathbb R)$, and 
$\Delta_{\rm d}\in C^\infty(\mathbb R\times\mathcal U,\mathbb R^{1\times\rho})$ is the decaying error, satisfying $\Delta_{\rm d} (t,\mu)\to 0(t\to \infty)$. If the system is $T$-periodic, then $R(\cdot,\cdot),R_0(\cdot)$ and $R_k(\cdot)$ are all required to be $T$-periodic with $\Delta_{\rm d} (t,\mu)\equiv 0$.

When~\eqref{eq-flp} holds, such an IM can be designed as:
\begin{equation}
    \begin{aligned}
        \Xi(t) &= I_{N_R+1} \otimes S(t), \\
        \Gamma(t) &= \begin{bmatrix}
            R_0(t) & R_1(t) & \cdots & R_{N_R}(t)
        \end{bmatrix}.
    \end{aligned} \label{eq-rim}
\end{equation}
\end{theorem}

\begin{proof}
  See~\ref{pf-thm-flp} for a proof.
\end{proof}

\begin{corollary}[Regulation via the IM Mechanism]\label{cr-flp}
Assuming the conditions in Theorem~\ref{thm-flp}, suppose a regulator of the form~\eqref{controller} has stabilized the closed-loop system.
If it embeds an asymptotic IM for $\{S(\cdot),R(\cdot,\cdot)\}$, then it solves  Problem~\ref{prob-1}.
Conversely, suppose the regulator solves Problem~\ref{prob-1} through the zero-input IM mechanism.
This means the forced steady-state controller output in~\eqref{REiff2} is asymptotically reproducible by the embedded zero-input IM; namely,
there exists a UB mapping $\bar \Sigma(\cdot,\cdot)$, induced by the embedded IM and satisfying~\eqref{RE3}
such that $H(t)[\Sigma(t,\mu)-\bar \Sigma(t,\mu)]\to 0 (t\to \infty)$,
where $\Sigma(\cdot,\cdot)$ is a UB solution to~\eqref{REiff2}.
Then the embedded IM is an asymptotic IM for $\{S(\cdot),R(\cdot,\cdot)\}$.

Consequently, provided the stabilizing realization in~\ref{sec-st} is feasible, Problem~\ref{prob-1} is solvable through a linear finite-dimensional zero-input IM mechanism if and only if the finite linear parameterization~\eqref{eq-flp} holds.
\end{corollary}

\begin{proof}
  See~\ref{pf-cr-flp} for a proof.
\end{proof}

\begin{remark}\label{rm-flp-explain}
  Theorem~\ref{thm-flp} reveals the exact algebraic boundary~\eqref{eq-flp}  for a linear finite-dimensional IM. 
  It can accommodate uncertainty only when the non-decaying component of the error-zeroing mapping $R(t,\mu)$ belongs to a finite-dimensional function family, with all parameter dependence carried by finitely many scalar coefficients~$\theta_{R k}(\mu)$.
  This aligns with the observation that the output trajectories of the unforced system~\eqref{eq-isi-im} always reside in a finite-dimensional function family.
  Notably, marginal stability of the generator is not used in the necessity proof of~\eqref{eq-flp}.
\end{remark}

\begin{remark}
  Corollary~\ref{cr-flp} explains the zero-input IM mechanism represented by the unforced system~\eqref{eq-isi-im}, or~\eqref{eq-imdef1}, which is the classical signal-generation viewpoint adopted in LTI, periodic and nonlinear output regulation. 
  Regulators operating through this mechanism are referred to in this paper as \textit{internal-model-based}  regulators.
  In the periodic case, Proposition~\ref{pr-re} shows that the strengthened equations~\eqref{RE3}-\eqref{RE4} are necessary and sufficient.
  Hence, restricting attention to the zero-input IM mechanism incurs no additional loss, and~\eqref{eq-flp} gives the \textit{exact solvability boundary} for finite-dimensional periodic robust regulation.
  For general LTV systems, however, only the weaker equations~\eqref{REiff2} are necessary.
  The vanishing error-driven term $G(\cdot)(C(\cdot,\cdot)\Pi(\cdot,\cdot)+Q(\cdot,\cdot))$ may produce a persistent contribution to the control input that is not asymptotically reproducible by the zero-input IM.
  Therefore, the nonexistence of an asymptotic IM rules out exact regulation through the zero-input IM mechanism, but does not by itself exclude regulators exploiting this more general forced mechanism, which may be studied in the future.
\end{remark}

\begin{remark}\label{rm-flp-compare-ti}
  In the LTI case, \eqref{eq-flp} is automatically satisfied, because $R(\mu)$ is a constant and can always be expanded on the standard bases.
  Indeed, the LTI robust design  based on Cayley-Hamilton theorem~\cite[Sec. 1.4]{isidori2003robust} can be regarded as taking advantage of this property, where the robust IM only cares about the size of $R(\mu)$.
  However, the LTV robustness is different by nature, because  $R(t,\mu)$ is a function of $t$ and may even constitute an infinite-dimensional function family, leading to the failure of~\eqref{eq-flp} as shown in Section~\ref{sec:gid}.
  Besides, the friend-centric perspective for nonlinear IM designs~\cite{bin2016robust,astolfi2022harmonic} assumes uncertainties influence $R(\cdot,\cdot)w$ within a known class of signals, and~\eqref{eq-flp} indicates for the LTV case,  this assumption characterizes the limit of what a finite-dimensional IM can accommodate.
\end{remark}

\begin{remark}\label{rm-isi-compare-marconi}
  For linear systems with periodic state jumps,  \cite[Sec. IV-C]{marconi2013internal} put forward the following assumption for robust IMs, namely 
  \begin{equation}
    R(t,\mu) = \tilde R (t)\gamma (\mu) ,\nonumber
  \end{equation}
  with $\tilde R \in C^\infty(\mathbb{R}, \mathbb{R}^{1\times \tilde r})$ known and $\gamma\in C^\infty(\mathcal{U},\mathbb{R}^{\tilde r\times \rho})$ uncertain.
  This means the influence of $\mu$ within $R(t,\mu)$ should be separated, which is equivalent to the finite linear parameterization~\eqref{eq-flp} in periodic systems (See Supplementary~S2).

  The Cayley-Hamilton-like condition in~\cite[Lem 3.2]{zhang2009tac} can likewise be interpreted as a checkable sufficient condition for~\eqref{eq-flp}.
  It assumes there exists an integer~$q$ and $T$-periodic functions~$a_0(t),a_1(t),\cdots,a_{q-1}(t)$ such that
  \begin{equation}
    L_S^q R(t,\mu) +\sum_{i=0}^{q-1}a_i(t)L_S^i R(t,\mu) =0 \nonumber
  \end{equation}
  for all $t\in [0,T)$ and all $\mu\in \mathcal{P}$. 
  Because this is a linear homogeneous ordinary differential equation with all its coefficients independent of $\mu$, its fundamental solution space is finite-dimensional with bases $\mu$-independent. 
  Denote the bases as $R_k(t)$ and their combination coefficients after extracting $R_0(t)$ as $\theta_{R k}(\mu)$, yielding~\eqref{eq-flp}.
\end{remark}

\subsection{Dimension Explosion: The General Consequence for Robust Regulation}  \label{sec:gid}
After characterizing the error-zeroing families that can be reproduced by a linear finite-dimensional IM,
this section investigates the family actually generated by the forced system~\eqref{eq-isi2}.
If uncertainty already makes the forcing or output terms in~\eqref{eq-isi2} span an infinite-dimensional function space, such complexity may be inherited directly by $R(\cdot,\cdot)$. 
Hence, the more revealing question is whether dimension explosion can still occur when the uncertain system matrices themselves belong to a finite-dimensional affine family defined in Definition~\ref{def-apu}, and Example~\ref{eg-inf} answers this question affirmatively. 
Here, \emph{dimension explosion} refers to the phenomenon that the non-decaying component of $\{R(\cdot,\mu):\mu\in\mathcal P\}$ spans an infinite-dimensional function space, despite the finite-dimensionality of the underlying parametric uncertainty.

\begin{definition}[Affine Parametric Uncertainty] \label{def-apu}
A matrix-valued function $\Theta(t,\mu)$ is said to be subject to affine parametric uncertainty if 
\begin{equation}
\Theta(t,\mu)=
\Theta_0(t)+\sum_{i=1}^{N_\Theta}
\theta_{\Theta i}(\mu) E_{\Theta i}(t),
\label{eq-affine-unc}
\end{equation}
where $\Theta_0(t)=\Theta(t,\mathbf{0})$ is the known nominal matrix, $E_{\Theta i}(t)$ is the known perturbation channel, and $\theta_{\Theta i}\in C^{\infty}(\mathcal{U},\mathbb{R})$ is a scalar coefficient function. The functions $E_{\Theta i}(\cdot)$ are smooth and UB, together with the derivatives required in the sequel. The coefficient functions $\theta_{\Theta i}(\cdot)$  satisfy
$\theta_{\Theta i}(\mathbf{0})=0$.
To make the representation nondegenerate, let
$E_{\Theta i}(\cdot)\not\equiv \mathbf{0}$,
and  $  \theta_{\Theta i}(\cdot) \not\equiv 0$ on $ \mathcal P$.
\end{definition}
Affine parametric uncertainty is finite-dimensional in the sense that all perturbations $(\Theta(t,\mu)-\Theta_0(t))$ belong to the finite-dimensional span generated by $E_{\Theta i}(t)$. Here, ``affine'' refers to the dependence on the scalar coefficients $\theta_{\Theta i}(\mu)$.
It automatically holds when matrices are constants. For example, $A(\mu)$ can always be expanded on the standard bases, thus satisfying~\eqref{eq-affine-unc}.

\begin{example}[Motivating Example]\label{eg-inf}
Consider a simple periodic system of the form~\eqref{sys}, where 
\begin{equation}
  \begin{aligned}
    &A = 0,\, B=1,\, C(t,\mu)=\sin(t) +2+\mu,\\
    &S=0,\,Q=1,\,P=0,\mu\in [-0.5,0.5],
  \end{aligned} \nonumber
\end{equation}
 which satisfies Assumptions~\ref{ass-s}-\ref{ass-m} with $r = 1$.
Calculations yield $b(t,\mu) = C(t,\mu)$,
%$\, b(t,\mu) = \mathscr{O}_A (t,\mu) =C(t,\mu),\,\bar {\mathscr{O}}_A(t,\mu) = \dot C(t,\mu),\,\mathscr{O}_S = 1,\,\bar {\mathscr{O}}_{S} = \mathscr{P} = \bar {\mathscr{P}}=0$.
%$M(t,\mu) = -\dot C(t,\mu)/C(t,\mu),\, N = 0,\,$
$\Phi_M(t,t_0,\mu) = C(t_0,\mu)/C(t,\mu)$,
%\, \Phi_S = 1$ 
$\Pi(t_0,\mu)= -1/C(t_0,\mu)$,
and 
$  R(t,\mu) = \dot C(t,\mu)/C^2(t,\mu)=\cos(t)/(\sin(t)+2+\mu)^2$.
Note the family 
\begin{equation}
  \mathcal{F}_{\rm e} = \left\{
  R_{\mu}(t)=\frac{\cos(t)}{(\sin(t)+2+\mu)^2}\,\middle|\,
  \mu \in [-0.5,0.5]
  \right\}   \nonumber
\end{equation}
is infinite-dimensional (See Supplementary~S3), but~\eqref{eq-flp} is a finite-dimensional function family, meaning $\mathcal{F}_{\rm e}$ cannot admit a  finite linear parameterization. According to Corollary~\ref{cr-flp}, Problem~\ref{prob-1} cannot be solved by a finite-dimensional IM-based regulator.
\end{example}

Example~\ref{eg-inf} shows that finite-dimensional affine parametric uncertainty can already generate an infinite-dimensional error-zeroing family. 
As revealed by~\eqref{eq-isi2}, the two principal mechanisms are the inversion of the high-frequency gain~$b(\cdot,\cdot)$ and the uncertainty-dependent transition matrix of $M(\cdot,\cdot)$, or the zero-dynamics~$A_\eta(\cdot,\cdot)$ when  adopting Byrnes-Isidori form. 
These observations are formalized by Lemmas~\ref{lm-infb}-\ref{lm-inf2}, which use one-channel affine uncertainties to isolate the minimal mechanisms. Their conclusions extend beyond the affine setting to any uncertainty family containing the corresponding one-parameter affine subfamily.

\begin{lemma}\label{lm-infb}
  With Assumptions~\ref{ass-s}-\ref{ass-m}, consider Problem~\ref{prob-1} in Byrnes-Isidori form,  where $b(t,\mu)$ is subject to one-channel affine parametric uncertainty, namely
  \begin{equation}
    b(t,\mu)=b_0(t)+\bar \mu E_b(t),\qquad \bar \mu\in\mathcal I_b, \nonumber
  \end{equation}
  with $\bar \mu=\theta_b(\mu)$ and $\mathcal I_b=\theta_b(\mathcal P)\subset\mathbb R$ as a nondegenerate interval.
  Define $\bar E_b (t)=E_b(t)/b_0(t)$ and assume the first $k$ derivatives of the error-zeroing matrix $R(\cdot,\cdot)$ are UB with $k=\max\{1,r-2\}$.

  If there exist $v_b\in\mathbb R^\rho$, $\phi_R>0$, a nondegenerate interval $\bar {\mathcal I}\subset\mathbb R$, and a sequence of intervals $\mathcal{J}_\ell=[a_\ell,c_\ell]$, $\ell\ge1$, with $a_\ell\to+\infty$, such that, for every $\ell$, $|\bar R_0(t)v_b|\geq \phi_R$ for all $t\in \mathcal{J}_\ell$, and $\bar {\mathcal I}\subset \bar E_b ( \mathcal{J}_\ell)$,
  then Problem~\ref{prob-1} cannot be solved by a linear finite-dimensional IM-based regulator.
\end{lemma}
\begin{proof}
  See~\ref{pf-lm-infb} for a proof.
\end{proof}

\begin{remark}\label{rm-infb}
    Lemma~\ref{lm-infb} shows that the inversion of even a single time-varying affine uncertainty channel may readily generate an infinite-dimensional family of non-decaying error-zeroing matrices.
    Its conditions  mean the uncertainty-induced variation of  $b^{-1}(t,\mu)$ remains both asymptotically visible and asymptotically rich, with a simple example as $b_0 = 1, E_b(t) = \sin(t)$ and $\bar R_0$ a nonzero constant.
    This dimension explosion is avoided when  $E_{bi}(\cdot)/b_0(\cdot)$ converges to constants for every $i$, because then $b^{-1}(t,\mu) \to (1+\bar \theta(\mu))b_0^{-1}(t)$ for some $\bar \theta\in C^\infty(\mathcal{U},\mathbb{R})$, thus making $R(\cdot,\cdot)$ satisfy~\eqref{eq-flp}.
    In particular, this automatically holds when $b(\mu)$ is time-invariant.
\end{remark}

\begin{lemma}\label{lm-inf}
  With Assumptions~\ref{ass-s}-\ref{ass-m}, consider a $T$-periodic Problem~\ref{prob-1} in Byrnes-Isidori form with scalar zero dynamics, namely $r=n-1$.
  Suppose that $A_\eta(t,\mu)$ is subject to constant one-channel affine parametric uncertainty, namely
  \begin{equation}
    A_\eta(t,\mu)=A_{\eta0}(t)+\bar \mu, \quad \bar \mu \in \mathcal I_\eta, \nonumber
  \end{equation}
  with $\bar \mu=\theta_\eta(\mu)E_\eta$, and $\mathcal I_\eta=\theta_\eta(\mathcal P)E_\eta\subset\mathbb R$ as a  nondegenerate interval.
  Floquet Theorem indicates that $\Phi_{\eta0}(t,\tau)=T_\eta(t)e^{\lambda_\eta(t-\tau)}T_\eta^{-1}(\tau)$ with $T_\eta(t)$ nonzero and $T$-periodic; $\Phi_S(t,\tau) =T_S(t)e^{S_0(t-\tau)}T_S^{-1}(\tau)$ with $T_S(t)$ invertible and $T$-periodic.
  Define the equivalent periodic input 
  \begin{equation}
  U(t)=T_\eta^{-1}(t)\big(P_{\mathrm{BI},l}(t)-\beta(t)Q(t)\big)T_S(t) =  \sum_{k\in\mathbb Z}U_{k}e^{\mathrm j k\omega t}, \nonumber
  \end{equation}
  which admits the above Fourier expansion with $\omega=2\pi/T$.
  Partition $\alpha^T(t)=\begin{bmatrix}\alpha_u^T & \alpha_\eta(t)\end{bmatrix}$ with $\alpha_\eta(t)$ a scalar.

  If $U(t)$ contains infinitely many nonzero Fourier coefficients $U_{k}$, and $\alpha_\eta(t)$ does not vanish identically on any nondegenerate subinterval of [0,T], 
  then Problem~\ref{prob-1} cannot be solved by a linear finite-dimensional IM-based regulator.
\end{lemma}

\begin{proof}
  See~\ref{pf-lm-inf} for a proof.
\end{proof}

\begin{remark}\label{rm-inf}
Lemma~\ref{lm-inf} shows that even a constant perturbation of a single zero-dynamics channel produces parameter-dependent gains at infinitely many excited frequencies. 
As $\bar \mu$ varies, the resulting gain profiles cannot be represented by finitely many $\bar \mu$-independent frequency patterns, so the corresponding steady-state responses span an infinite-dimensional family beyond the capability of a finite-dimensional IM. 
By contrast, an LTI regulation problem contains only the finitely many modes of the exosystem, which can always be reproduced by a finite-dimensional IM. 
This is a frequency-domain interpretation of the failure of~\eqref{eq-flp} for LTV systems, thus complementing the algebraic viewpoint in Remark~\ref{rm-flp-compare-ti}.

The requirement that $U(\cdot)$ have infinitely many nonzero Fourier coefficients is natural for a general periodic signal, while the condition on $\alpha_\eta(\cdot)$ only ensures that the resulting zero-dynamics variation is visible in $R(\cdot,\cdot)$.
Compared to Remark~\ref{rm-infb}, these conditions may hold even when the plant is LTI.
For example, let
$S=\beta=Q=\mathbf{0},A_\eta(\mu)=-2+\bar \mu ,b=\alpha_\eta=1,P_{\mathrm BI,l}(t)=e^{\sin(\omega t)}$,
with $\mathcal I_\eta\subset(-\infty,1)$.
Then $U(t)=e^{\sin(\omega t)}$ has infinitely many nonzero Fourier coefficients, and Lemma~\ref{lm-inf}
applies. 
The same conclusion extends to multi-dimensional zero dynamics whenever they contain an uncertain Floquet mode that is both excited and visible.
\end{remark}

\begin{lemma}\label{lm-inf2} 
  With Assumptions~\ref{ass-s}-\ref{ass-m}, consider a $T$-periodic Problem~\ref{prob-1} in Byrnes-Isidori form with scalar zero dynamics, namely $r=n-1$. Suppose that $A_\eta(t,\mu)$ is subject to one-channel affine parametric uncertainty, namely 
  \begin{equation} 
    A_\eta(t,\mu)=A_{\eta0}(t)+\bar \mu E_\eta(t),\qquad \bar \mu \in\mathcal I_\eta,\nonumber
\end{equation}
with $\bar \mu =\theta_\eta(\mu)$, $\mathcal I_\eta=\theta_\eta(\mathcal P)\subset\mathbb R$ as a  nondegenerate interval, and  $E_\eta(\cdot)$ $T$-periodic.
The Floquet representations for $\Phi_{\eta0}(t,\tau)$ and $\Phi_{S}(t,\tau) $, the definition of $U(t)$, and the partition of  $\alpha^T(t)$ are the same as those in Lemma~\ref{lm-inf}.

If there exists $t_*\in [0,T)$ such that
$E_\eta(t_*)=0$,
$\dot E_\eta(t_*)\neq0$,
$U(t_*)\neq0$,
$\alpha_\eta(t_*)\neq0$,
then Problem~\ref{prob-1} cannot be solved by a linear finite-dimensional IM-based regulator.
\end{lemma}

\begin{proof}
  See~\ref{pf-lm-inf2} for a proof.
\end{proof}

\begin{remark}\label{rm-inf2}
Lemma~\ref{lm-inf2} complements Lemma~\ref{lm-inf} by revealing a different mechanism of dimension explosion. While Lemma~\ref{lm-inf} relies on infinitely many excited frequencies in $U(\cdot)$, the present result imposes no spectral richness condition: even a finite-spectrum, or constant, equivalent input may be transformed by a periodically uncertain zero-dynamics channel into an error-zeroing family that is infinite-dimensional. The conditions are readily satisfied; for example, one may take $U(t)\equiv U_0\neq \mathbf{0}$ and $E_\eta(t)=\sin(\omega t)$, provided $\alpha_\eta(0)\neq0$. Thus, either a rich equivalent input or a time-varying uncertainty channel can make zero-dynamics uncertainty generate an infinite-dimensional family.
\end{remark}

Consequently,  Example~\ref{eg-inf} and Lemmas~\ref{lm-infb}-\ref{lm-inf2} independently imply the following intrinsic deficit of finite-dimensional IM-based regulators.
\begin{theorem}\label{thm-fail}
  There exist problems of LTV robust regulation in the form of  Problem~\ref{prob-1}, which cannot be solved by any linear finite-dimensional internal-model-based regulator.
\end{theorem}

\begin{remark}\label{rm-inf-conc}
The results here show that the failure of linear finite-dimensional IMs is already present under highly structured one-parameter uncertainty, and therefore cannot be attributed merely to the breadth of the admissible uncertainty class. 
This clarifies why merely embedding the exosystem dynamics, as in the classical LTI internal model principle, is generally insufficient for uncertain LTV systems: 
the IM must also reproduce the plant-dependent deformation of the exogenous signals, which may cause dimension explosion. 
A related impossibility result for nonlinear systems is established in~\cite{bin2022robustness} for general smooth finite-dimensional regulators under arbitrarily small unstructured $C^0$ perturbations. Whereas the signal complexity there is supplied directly by the richness of the perturbation class, here it is dynamically generated from highly structured one-parameter uncertainty through the time-varying RE. 
This shows that the obstruction can be rooted in the time-varying system structure itself, making~\eqref{eq-flp} a genuinely restrictive condition for uncertain LTV systems.
\end{remark}

\section{Robust Regulator Designs} \label{sec-rb}

  Section~\ref{sec-ru} identifies uncertainty classes ensuring finite linear parameterization a priori and hence admitting exact finite-dimensional regulation.
  Then Section~\ref{sec-su} develops finite-dimensional approximate IMs for more general analytic uncertainties.
  
\subsection{Exact Regulation for Regular Uncertainty} \label{sec-ru}

  The uncertainty that ensures finite linear parameterization for $R(\cdot,\cdot)$ is termed \textit{regular uncertainty}.
  Two specific cases are provided by identifying how the uncertainty enters the system.
  Considering~\eqref{eq-isi2}, when only $\{P(\cdot,\cdot),Q(\cdot,\cdot)\}$, or $ \{\alpha(\cdot,\cdot) , \beta(\cdot,\cdot) , P_{\rm BI}(\cdot,\cdot), Q(\cdot,\cdot)\}$ in Byrnes-Isidori form, are subject to affine parametric uncertainty, the affine dependence is preserved for $R(\cdot,\cdot)$, as shown in Theorems~\ref{thm-rbpq}-\ref{thm-rbfd}. 

\begin{theorem}\label{thm-rbpq}
  With Assumptions~\ref{ass-s}-\ref{ass-m}, consider Problem~\ref{prob-1},  where only   $\{P(t,\mu)$,$Q(t,\mu)\}$ are subject to affine parametric uncertainty. 
Then there exists an asymptotic IM  as  
\begin{equation}
  \begin{aligned}
      \Xi(t) &= I_{(N_P+N_Q+1)} \otimes S(t),\\
    \Gamma(t)&=  \begin{bmatrix}
   \bar  R_0(t)  &\bar  R_{P}(t) &\bar  R_{Q}(t)
   \end{bmatrix}, \\
  \bar R_{P}(t) &=
   \begin{bmatrix}
  \bar   R_{P1}(t)&\bar R_{P2}(t)&\cdots&\bar  R_{PN_P}(t) 
   \end{bmatrix} ,\\
  \bar R_{Q}(t) &=
   \begin{bmatrix}
   \bar   R_{Q1}(t)&\bar R_{Q2}(t)&\cdots&\bar  R_{QN_Q}(t) 
   \end{bmatrix} ,
  \end{aligned} \label{eq-sim-pq}
\end{equation}
where $\bar R_0(t)$ is the nominal error-zeroing matrix;
 $\bar R_{Pi}(t)$ and $\bar R_{Qi}(t)$ are calculated according to Theorem~\ref{thm-re}
  by replacing $\{P(\cdot,\cdot),Q(\cdot,\cdot)\}$ with $\{E_{Pi}(t),\mathbf{0}\}$ and $\{\mathbf{0},E_{Qi}(t)\}$, respectively.
\end{theorem}
\begin{proof}
   See~\ref{pf-thm-rbpq} for a proof.
\end{proof}

\begin{remark}
  Theorem~\ref{thm-rbpq} takes advantage of the linearity of $\bar R(\cdot,\cdot)$ in $\{P(\cdot,\cdot),Q(\cdot,\cdot)\}$, and the derivation is coordinate-free.
  It designs sub-IMs for the nominal system and each uncertainty channel, which are then combined as \eqref{eq-sim-pq}. 
\end{remark}

\begin{theorem}\label{thm-rbfd}
   With Assumptions~\ref{ass-s}-\ref{ass-m}, consider Problem~\ref{prob-1} in Byrnes-Isidori form,  where only $ \{\alpha(t,\mu) , \beta(t,\mu) , P_{\rm BI}(t,\mu), Q(t,\mu)\}$  are subject to affine parametric uncertainty. 
  Then there always exists a finite-dimensional asymptotic IM.
\end{theorem}
\begin{proof}
  The proof follows that of Theorem~\ref{thm-rbpq}, with only a few hints given here. There may exist cross terms like $\theta_{\beta 1}(\mu) \theta_{Q 2}(\mu)$, which can be regarded as new scalar coefficient functions of $\mu$, thus preserving finite linear parameterization. 
\end{proof}

\begin{remark}
  Theorems~\ref{thm-rbpq}-\ref{thm-rbfd} provide directly checkable plant-level conditions for finite-dimensional IM realizability.
  By comparison, the conditions in~\cite{zhang2009tac,marconi2013internal} are formulated in terms of structural properties of the resulting  $R(\cdot,\cdot)$.
\end{remark}

\subsection{Approximate Regulation Beyond Regular Uncertainty}\label{sec-su}

  The uncertainty that invalidates the finite linear parameterization condition for $R(\cdot,\cdot)$ is defined as \textit{singular uncertainty}.
  For such uncertainty, Corollary~\ref{cr-flp} precludes exact robust regulation by a finite-dimensional IM-based regulator.
  This limitation motivates us to relax the control objective from exact regulation to approximate regulation.
  Indeed, the design developed in this section applies to general analytic parametric uncertainty, not restricted to singular uncertainty or affine parametric uncertainty.

  The derivations below focus on the two principal sources of dimension explosion revealed in Section~\ref{sec:gid}, namely $b^{-1}(\cdot,\cdot)$ and $\Phi_\eta(\cdot,\cdot,\cdot)$. 
  Uncertainties in other matrices can be treated similarly and introduce no additional conceptual difficulty. 
  In order to approximate the linear parameterization, we expand $b^{-1}(\cdot,\cdot)$  and $\Phi_\eta(\cdot,\cdot,\cdot)$ into parameterized series and truncate them to obtain a finite-dimensional approximate IM with a bounded residual.

\begin{proposition}[Analytic Approximation] \label{pr-app}  
With Assumptions~\ref{ass-r}-\ref{ass-m}, suppose the high-frequency gain $b(t,\mu)$ and the zero-dynamics matrix $A_\eta(t,\mu)$ are uniformly real-analytic functions of $\mu \in \mathcal{U}$.
If
\begin{equation}
   \phi_\mu < \frac{1}{C_b },\quad \phi_\mu < \frac{1}{C_\eta }, \label{eq-conds}
\end{equation}
where $C_b,C_\eta>0$ are positive constants determined by $b(\cdot,\cdot)$ and $A_\eta(\cdot,\cdot)$ (See Supplementary~S11),
then the $k_b$-th order  approximation for $b^{-1}(\cdot,\cdot)$ is
\begin{equation}
    b^{-1}(t,\mu) = b_0^{-1}(t) + \sum_{i=1}^{\bar N_b} \bar {\theta}_{bi}(\mu) b_{i}^{-1}(t) + \Delta_b^{k_b}(t,\mu), \label{eq-tr-b}
\end{equation}
with $\bar N_b = \binom{N+k_b}{N}-1$, and the $k_\eta$-th order approximation for $\Phi_\eta(\cdot,\cdot,\cdot)$ is
\begin{equation}
  \begin{aligned}
  \Phi_\eta(t,t_0,\mu) =& \Phi_{\eta_0}(t,t_0) + \sum_{i=1}^{\bar N_\eta} \bar \theta_{\eta i}(\mu) \Phi_{\eta i}(t,t_0)
   + \Delta_{\eta}^{k_\eta}(t,t_0,\mu),
    \end{aligned}\label{eq-tr-eta}
\end{equation}
with $\bar N_\eta = \binom{N+k_\eta}{N}-1$.
The truncation errors satisfy
\begin{equation}\label{eq-bds}
  \begin{aligned}
       |\Delta_b^{k_b}(t,\mu)| &\le M_b(\bar{C}_b \phi_{\mu})^{k_b+1}, \\
    ||\Delta_\eta^{k_\eta}(t,t_0,\mu)|| &\le M_\eta e^{-\bar \phi_\eta(t-t_0)} (\bar{C}_\eta \phi_{\mu})^{k_\eta+1},
  \end{aligned}
\end{equation}
where $M_b,M_\eta > 0$ are constants independent of $k_b$ and  $k_\eta$; $\bar C_b,\bar \phi_\eta,\bar C_\eta$ are constants satisfying $C_b < \bar{C}_b < \phi_\mu^{-1}$, $0<\bar \phi_\eta<\hat \phi_\eta$,
and  $C_\eta < \bar{C}_\eta < \phi_\mu^{-1}$.

%$\bar {\theta}_{bi}(\mu)$ and $b_{i}^{-1}(t)$ are defined as following.
For each multi-index $m = (m_1, \dots, m_N) \in \mathbb{N}_0^N$ ($1 \le |m| \le k_b$)  (See Supplementary~S1) mapped to the scalar index $i \in \{1, \dots, \bar N_b\}$, define:
\begin{equation}
    \bar {\theta}_{bi}(\mu) = \frac{1}{C_b^{-|m|}} \mu^m, b_{i}^{-1}(t) =C_b^{-|m|}  \left( \frac{1}{m!} D_\mu^m (b^{-1})(t,\mathbf{0}) \right), \nonumber
\end{equation}
where $b_i^{-1}(\cdot)$ is UB, and $C_b^{-|m|}$ is a scaling factor to normalize $b_i^{-1}(\cdot)$.
In particular, $\bar \theta_{b0}=1$ and $b_{0}^{-1}(t) = b^{-1}(t,\mathbf{0})$.
$\bar \theta_{\eta i}(\mu)$ and $\Phi_{\eta i}(t,t_0)$ are defined analogously, by replacing $b^{-1}(t,\mathbf{0})$ with $\Phi_\eta(t,t_0,\mathbf{0})$, and the scaling factor is chosen as $\tilde C_\eta^{-|m|}$ with $\tilde C_\eta>C_\eta$ large enough (See Proposition~\ref{pr-bd}). Besides, $\Phi_{\eta i}(t,t_0)$ is uniformly exponentially decaying as $(t-t_0)\to \infty$.
\end{proposition}

\begin{proof}  
See~\ref{pf-pr-app} for a proof.
\end{proof}

\begin{lemma}[Approximate Internal Model]\label{lm-im-app}
  With Assumptions~\ref{ass-s}-\ref{ass-m}, consider Problem~\ref{prob-1} in Byrnes-Isidori form,  where $ \{A_\eta(t,\mu) , b(t,\mu) \}$  are uniformly real-analytic functions of $\mu \in \mathcal{U}$.
  If~\eqref{eq-conds} holds,
  there exists an approximate IM shown as~\eqref{eq-sim-app}, with the approximation error $\Delta^{k_b,k_\eta}_R (t,\mu)$ satisfying
  \begin{equation} \label{eq-err-app}
      \left|\left|\Delta^{k_b,k_\eta}_R (t,\mu)\right|\right| 
       <\phi_0 \bigg(  \big(\bar{C}_b \phi_{\mu} \big) ^{k_b+1}
       +    \big(\bar{C}_\eta \phi_{\mu}\big)^{k_\eta+1} \bigg).
\end{equation}
\end{lemma}

\begin{proof}
The proof is similar to that of Theorem~\ref{thm-rbpq}.
Since only $A_\eta(\cdot,\cdot),b(\cdot,\cdot)$ are uncertain, $\bar {\mathscr{O}}_A(t),\mathscr{P}(t),\bar {\mathscr{P}}(t)$ are independent of $\mu$.
Substitute \eqref{eq-tr-eta} into \eqref{eq-pi-BI} to obtain
\begin{equation}
\bar \Pi(t,\mu) =\bar \Pi_0 (t) + \sum_{i=1}^{\bar N_\eta}\bar \theta_{\eta i}(\mu)  \bar\Pi_{i}(t)+\Delta^{k_\eta}_\Pi (t,\mu), \nonumber
\end{equation} 
where
$\bar \Pi_{i,{\rm u}}(t) =\mathbf{0}$,
$\bar \Pi_{i,{\rm l}}(t) =\int_{t_0}^{t} \Phi_{\eta i}(t,\tau) 
( P_{\rm  l}(\tau) - \beta(\tau) Q(\tau))\Phi_S(\tau,t)  \,\mathrm{d}\tau  $,
$\Delta^{k_\eta}_{\Pi,{\rm u}} (t,\mu)=\mathbf{0}$, and  
$
     \Delta^{k_\eta}_{\Pi,{\rm l}} (t,\mu) =
  \int_{t_0}^{t} \Delta_{\eta}^{k_\eta}(t,\tau,\mu)
   ( P_{\rm l}(\tau) - \beta(\tau) Q(\tau))
  \Phi_S(\tau,t)  \,\mathrm{d}\tau
$.
Substitute this with \eqref{eq-tr-b} into \eqref{eq-R}, and 
for each multi-index $m = (m_1, \dots, m_N) \in \mathbb{N}_0^N$ ($1 \le |m| \le k_b+k_\eta$)  mapped to the scalar index $i \in \{1, \dots, \bar N_R\}$ with $\bar N_R = \binom{N+k_b+k_\eta}{N}-1$,
denote $\mu^m$ with its scaling factors as $\theta_{Ri}(\mu)$, with the corresponding function as $\bar R_i(t)$.
This yields
\begin{equation}
 \bar R(t,\mu) =\bar R_0(t) + \sum_{i=1}^{\bar N_R} \theta_{Ri}(\mu)\bar R_i(t) +\Delta^{k_b,k_\eta}_R (t,\mu),
  \nonumber
\end{equation}
where  the approximation error
$
      \Delta^{k_b,k_\eta}_R (t,\mu) =- \Delta_{b}^{k_b}(t,\mu)[\bar {\mathscr{O}}_A(t)\bar\Pi(t,\mu)+\bar {\mathscr{O}}_S(t)+\bar {\mathscr{P}}(t)]  
      -[ b^{-1}(t,\mu) -\Delta_{b}^{k_b}(t,\mu) ]\bar {\mathscr{O}}_A(t) \Delta^{k_\eta}_\Pi (t,\mu) 
$
satisfies~\eqref{eq-err-app}.
Similar to Theorem~\ref{thm-flp}, the approximate IM is
\begin{equation}
    \begin{aligned}
    \Xi(t) &= I_{\bar N_R+1} \otimes S(t),\\
    \Gamma(t)&=  \begin{bmatrix}
   \bar R_0(t)  & \bar R_1(t) &\bar R_2(t) &\cdots &\bar R_{\bar N_R}(t)
   \end{bmatrix}.
  \end{aligned} \label{eq-sim-app}
\end{equation}

\end{proof}

\begin{remark}[Computational Realization]\label{rm-cal}
  The approximate IM~\eqref{eq-sim-app} can be calculated numerically without explicit solutions or calculating $\Phi_{\eta i}(\cdot,\cdot)$.  
 Applying the multi-index Leibniz rule  
 to~\eqref{eq-se-sim} at $\mu=\mathbf{0}$ yields that  for $|m|=0$,
 \begin{equation}
      \dot {\bar\Pi}_{0,\mathrm{l}}(t) = A_{\eta 0}(t) \bar \Pi_{0,\mathrm{l}}(t) -\bar\Pi_{0,\mathrm{l}}(t) S(t) +P_{\rm BI,l}(t) -\beta(t)  Q (t) , \nonumber
    \end{equation}
 and for $|m| \ge 1$, 
 let $\bar\Pi_{m,{\rm l}}(t) = D_\mu^m \bar \Pi_{\rm l}(t, \mathbf{0})$:
 \begin{equation}
  \begin{aligned}
     \dot{\bar \Pi}_{m,{\rm l}}(t) =& A_{\eta 0}(t)\bar\Pi_{m,{\rm l}}(t) - \bar\Pi_{m,{\rm l}}(t)S(t) 
     + \sum_{ k \le m,|k|>0} \binom{m}{k} D_\mu^k A_\eta(t,\mathbf{0})\bar \Pi_{m-k, {\rm l}}(t). \label{eq-compute-pi}
  \end{aligned}
 \end{equation}
The cascade ordinary differential equations can be integrated iteratively from $\bar \Pi_{0,{\rm l}}(t)$ with  all the initial values set as $\mathbf{0}$.
By uniqueness of the Taylor coefficients,
$\bar \Pi_{i,\mathrm{l}}(t) =\tilde C_\eta^{-|m|} \bar \Pi_{m,\rm{l}}(t)/m! $, where $\tilde C_\eta^{-|m|}$ and the mapping from $m$ to $i$ are defined in Proposition~\ref{pr-app}.
The remaining calculations are standard and are therefore omitted.
\end{remark}

\begin{remark} \label{rm-general-design}
  Lemma~\ref{lm-im-app} constructs the approximate IM via Taylor series of $b^{-1}(\cdot,\cdot)$ and $\Phi_\eta(\cdot,\cdot,\cdot)$, and a similar method can be applied directly to $R(\cdot,\cdot)$, if its expression is available, or to other uniformly real-analytic matrices like $\alpha(t,\mu)$.
  Consequently, the proposed approximate IM constitutes a general framework for approximate regulation.
  Besides, there are other approximation methods.
  For example, if the plant is LTI and the reference is $T$-periodic, one can decompose it into a Fourier series and retain finitely many components. The filtered reference can be generated by an LTI exosystem $\{\tilde S,\tilde P,\tilde Q\}$, reducing the problem to the LTI case.
\end{remark}

Because of the approximation error $\Delta_R(\cdot,\cdot)$, the final tracking error will not decay to zero but only be bounded, as shown in Theorem~\ref{thm-rbapp}.

\begin{theorem}[Approximate Regulation]\label{thm-rbapp}
    With Assumptions~\ref{ass-s}-\ref{ass-m}, consider Problem~\ref{prob-1} in Byrnes-Isidori form,  where $ \{A_\eta(t,\mu) , b(t,\mu),\dot b(t,\mu) \}$  are uniformly real-analytic functions of $\mu \in \mathcal{U}$. 
  %Assume the conditions in Proposition~\ref{pr-app}, namely~\eqref{eq-conds}.
  Suppose the approximate IM~\eqref{eq-sim-app} satisfies 
  \begin{equation}
    \int_{t-\phi_\delta}^{t} (\Gamma(\tau)\Phi_\Xi(\tau,t-\phi_\delta))^T \Gamma(\tau)\Phi_\Xi(\tau,t-\phi_\delta)   \,\mathrm{d}\tau  \geq \phi_{\rm uco} I\nonumber
  \end{equation}
  for constants $\phi_\delta(k_b,k_\eta), \phi_{\rm uco}(k_b,k_\eta) >0$, which may depend on the approximation orders.
Then  if $\phi_{\mu}  <\min\{C_b^{-1},C_\eta^{-1}\}$, the regulator~\eqref{eq-hg} constructed according to Propositions~\ref{pr-stable}-\ref{pr-cr} achieves approximate regulation, namely
\begin{equation}
  \begin{aligned}
 \limsup_{t\to\infty} &|e(t)| \leq
  \phi_0 \phi_w \bigg(  \big(\bar{C}_b \phi_{\mu} \big) ^{k_b+1}
       +    \big(\bar{C}_\eta \phi_{\mu}\big)^{k_\eta+1} \bigg) 
       \Bigg(\frac{e^{(\phi_S+a_3/a_2)\phi_\delta(k_b,k_\eta)}}{\phi_{\rm uco}(k_b,k_\eta)} (k_b+1)^N (k_\eta+1)^N\Bigg)^{2r+1},
    \end{aligned}
  \label{eq-app-e}
\end{equation}
where $\phi_w$ depends only  on the scale of $w$.
\end{theorem}
\begin{proof}
See~\ref{pf-thm-rbapp} for a proof.
\end{proof}

\begin{remark}
  The error bound in~\eqref{eq-app-e} is determined by a tradeoff between the approximation accuracy of the IM and its observability. 
  The canonical realization in Proposition~\ref{pr-cr} relies on the uniform complete observability of the approximate IM, and increasing $k_b$ and $k_\eta$ may increase $\phi_\delta(k_b,k_\eta)$ and/or decrease $\phi_{\mathrm{uco}}(k_b,k_\eta)$.
  This may enlarge $\phi_G$ and the sufficient controller gains~$k$ and $g$, thereby worsening the resulting worst-case bound on the closed-loop amplification of the IM approximation error.
  Consequently, as the approximation orders increase, the decay of $\Delta_R^{k_b,k_\eta}(\cdot,\cdot)$   alone does not necessarily imply a decreasing  tracking-error bound.
  In particular, if $\phi_\delta$ and $\phi_{\mathrm{uco}}^{-1}$ remain UB with respect to $k_b,k_\eta$ and $\phi_\mu$ is sufficiently small, then the error bound tends to zero as $k_b,k_\eta\to \infty$. 
\end{remark}

\section{Illustrative Example}
Consider a non-periodic LTV  problem, adapted from~\cite{zhang2009tac}, in which the angular displacement of a controlled pendulum is tuned according to that generated by another pendulum. The matrices of system~\eqref{sys} are given by
\begin{equation}
  \begin{aligned}
    S(t) &= \begin{bmatrix}
      0&1\\ -a - 2d (\cos(2t)+\cos(\sqrt{2}t))&0
    \end{bmatrix} , \\
    Q(\mu) & =\begin{bmatrix}
      -\bar \varepsilon(\mu) & -1
    \end{bmatrix} , \;P =\mathbf{0},\\
    A &= \begin{bmatrix}
      0&1\\ -a &0
    \end{bmatrix} , \;
    B = \begin{bmatrix}
      0\\ b
    \end{bmatrix} , \\
    C(\mu) &= \begin{bmatrix}
      \varepsilon(\mu) &1
    \end{bmatrix} ,
  \end{aligned} \nonumber
\end{equation}
where $a=1.6,b=1,d=0.1$; and $\varepsilon(\mu)\in [0.75,1.25]$, $\bar \varepsilon(\mu) \in [0.3,1.7]$ are uncertain parameters.
After transforming the plant into Byrnes-Isidori form and reparameterizing the uncertainties, we obtain
% T = [epsilon 1;1 0], hat x = T x
\begin{equation}
  \begin{aligned}
    Q(\mu) & =\begin{bmatrix}
      -1-\mu_2 & -1
    \end{bmatrix}, \;P_{\rm BI} =\mathbf{0},\\
    A_{\rm BI}(\mu) &= \begin{bmatrix}
      1+\mu_1&-a-(1+\mu_1)^2\\ 
      1 & -1-\mu_1
    \end{bmatrix},\;
    %A_{0} + \mu_1 E_{A1} +\mu_2 E_{A2}, \;
    B_{\rm BI} = \begin{bmatrix}
      b\\ 0
    \end{bmatrix} , \\
    C_{\rm BI} &= \begin{bmatrix}
      1 &0
    \end{bmatrix} ,
  \end{aligned} \nonumber
\end{equation}
%and  $ A_0 = [\varepsilon_0 , - a -\varepsilon_0^2 ; \beta ,A_{\eta 0}]$, $E_{A1} =[1 , 0; 0,E_\eta] $, $ E_{A2} = [ 0 , -1; 0,0 ]$
with $S(\cdot)$ unchanged, 
%where  $\beta = 1$,$A_{\eta 0} = -1$, $E_\eta = -1$,
$ \mu_1\in [-0.25,0.25]$,
and $ \mu_2 \in [-0.7,0.7]$.
The plant is minimum-phase with $r=1$.
We omit ``BI'' hereafter.
Define $\bar e(t)=e(t)/\phi_r$ as the normalized error, where $\phi_r$ is the peak magnitude of the reference $-Q(t,\mu)w(t)$. The normalized reference $\bar r(t) $ and the normalized output $\bar y(t)$ are defined analogously.
All simulations are performed in MATLAB$^{\rm TM}$ Simulink.
To evaluate robustness, the regular uncertainty and the singular uncertainty are tested separately,
with $w(0) = [0.5,-1]^T$, $L(0)=I $, and $d_0 =5$, throughout Examples~\ref{eg-iu}-\ref{eg-pu}.

\begin{example}\label{eg-iu}
  Consider the regular uncertainty  $\mu_2$ with $\mu_1=0$ fixed.
The nominal regulator is constructed using the nominal IM $\Xi_0(\cdot) =S(\cdot),\Gamma_0(\cdot) = R_0(\cdot)$,  
where $\Pi_0 = [1,1;1,0]$, $R_0(t) = -b^{-1} (CA_0 \Pi_0+Q_0S(t) )$.
The proposed robust regulator corresponds to $\Xi_{\rm r}(\cdot) =I_2 \otimes S(\cdot)  $,   $\Gamma_{\rm r}(\cdot) = [R_0(\cdot), R_Q(\cdot)]$,
where $\Pi_{Q,\mathrm{u}} = -E_Q $, and $\Pi_{Q,\mathrm{l}}(\cdot)$ is  calculated by integrating 
$ \dot {\Pi}_{Q,\mathrm{l}}(\cdot) = A_{\eta 0}{\Pi}_{Q,\mathrm{l}}(\cdot) - {\Pi}_{Q,\mathrm{l}}(\cdot) S(\cdot) - E_{Q}$, with $E_Q = [-1,0] $.
Then    $R_Q(\cdot) = -b^{-1}(CA_0\Pi_Q(\cdot)  + E_Q S(\cdot)  )$.
We follow the procedure in~\ref{sec-st}, and set  $\alpha_{\rm cr} = 4$, $k = 20,g=200$ for both regulators.
Set $\mu_2= 0.7$.
The simulation results are shown in Fig.~\ref{fig1}, where $\bar e_{\rm r}(t)$ converges to the level of machine precision,  thus verifying the robustness of the proposed regulator.
\begin{figure}[!t] 
\centerline{\includegraphics[width=0.9\columnwidth]{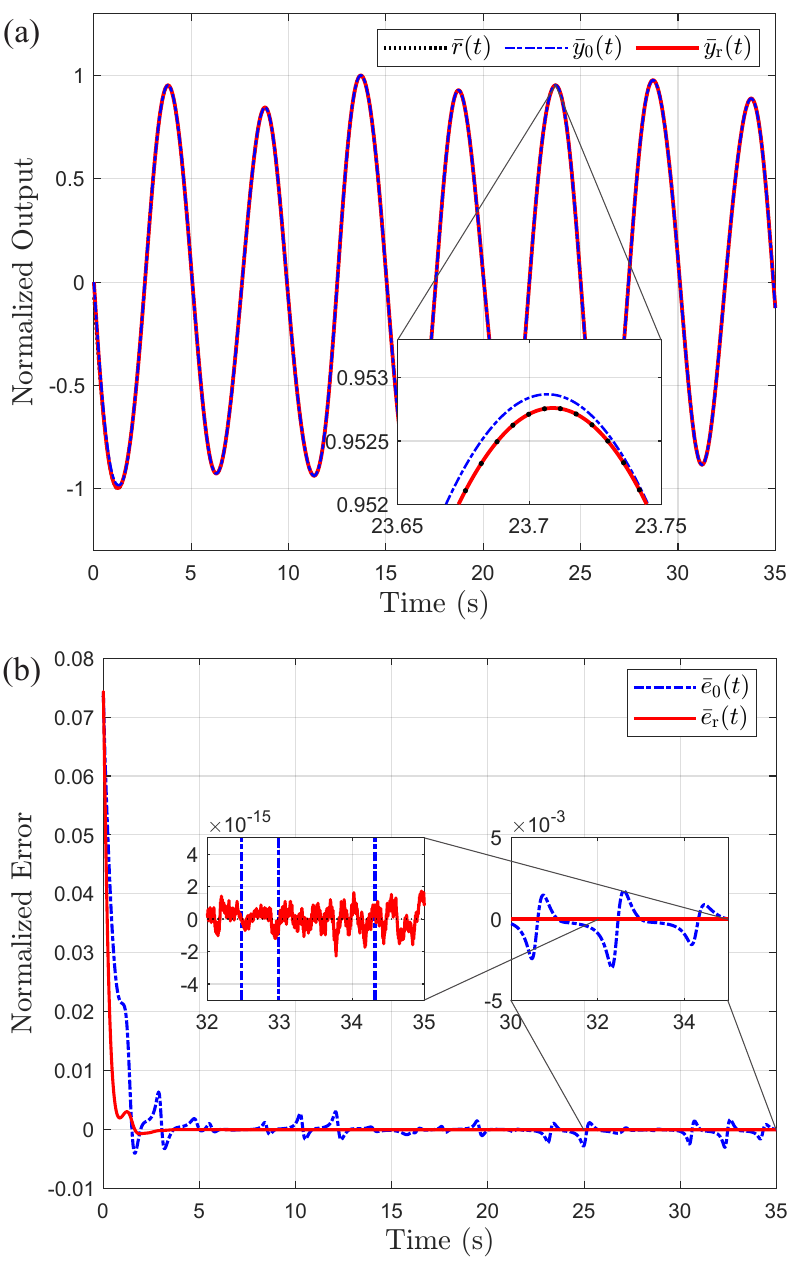}}
\caption{Simulation results of Example~\ref{eg-iu} for regular uncertainty, where $\bar r(t)$ denotes the normalized reference; $\bar y_0(t)$  and $\bar y_{\rm r}(t)$ denote the normalized outputs of the nominal regulator  and the robust regulator respectively;
$e_0(t)$ and $\bar e_{\rm r}(t)$ denote the normalized errors of the nominal regulator  and the robust regulator respectively.
}
\label{fig1}
\end{figure}    
\end{example}

\begin{example}\label{eg-pu}
  Consider the singular uncertainty $\mu_1$ with $\mu_2 = 0$ fixed, and Supplementary~S4 shows that $R(t,\mu_1)$ does not admit a finite linear parameterization, so no linear finite-dimensional IM-based regulator can achieve exact robust regulation. 
  Following Remark~\ref{rm-cal}, set $\tilde{C}_\eta=1$ and construct regulators of orders $0,1,\cdots,5$.
  The details are provided in Supplementary S4. 
  The $i$th-order regulator retains all terms through degree $i$ in $\mu_1$ in the approximation of $R(t,\mu_1)$. 
  Define 
  $\Delta \Gamma_0(t)=R_0(t) $, 
  $ \Delta \Gamma_1(t) = (a+1)\Pi_{1,\rm l}(t)+2\Pi_{0,\rm l}(t)+Q_0$,  and for 
  $i\geq2$, $ \Delta \Gamma_i(t)= (a+1)\Pi_{i,\rm l}(t) +2\Pi_{i-1,\rm l}(t) +\Pi_{i-2,\rm l}(t)$. 
  The IM of the $i$th-order regulator is therefore $\Xi_i(t)=I_{i+1}\otimes S(t)$, $ \Gamma_i(t)= [ \Delta \Gamma_0(t),\Delta \Gamma_1(t),\cdots,\Delta \Gamma_i(t) ]$. 
Set  $\alpha_{\rm cr} = 2$, $k=45,g=300$ for all the regulators to isolate the effect of the IM approximation from that of gain retuning.
% under a common stabilization setting.
Set $\mu_1= -0.25$, and the simulation results are shown in Fig.~\ref{fig2}, where  the asymptotic error level $\limsup_{t\to\infty}|\bar e_i(t)|$ is numerically estimated by $\max_{t\in[30,50]}|\bar e_i(t)|$.
They demonstrate that the proposed robust regulators achieve approximate regulation, with the residual normalized error decreasing rapidly as the approximation order increases. 

\begin{figure}[!t] 
\centerline{\includegraphics[width=0.9\columnwidth]{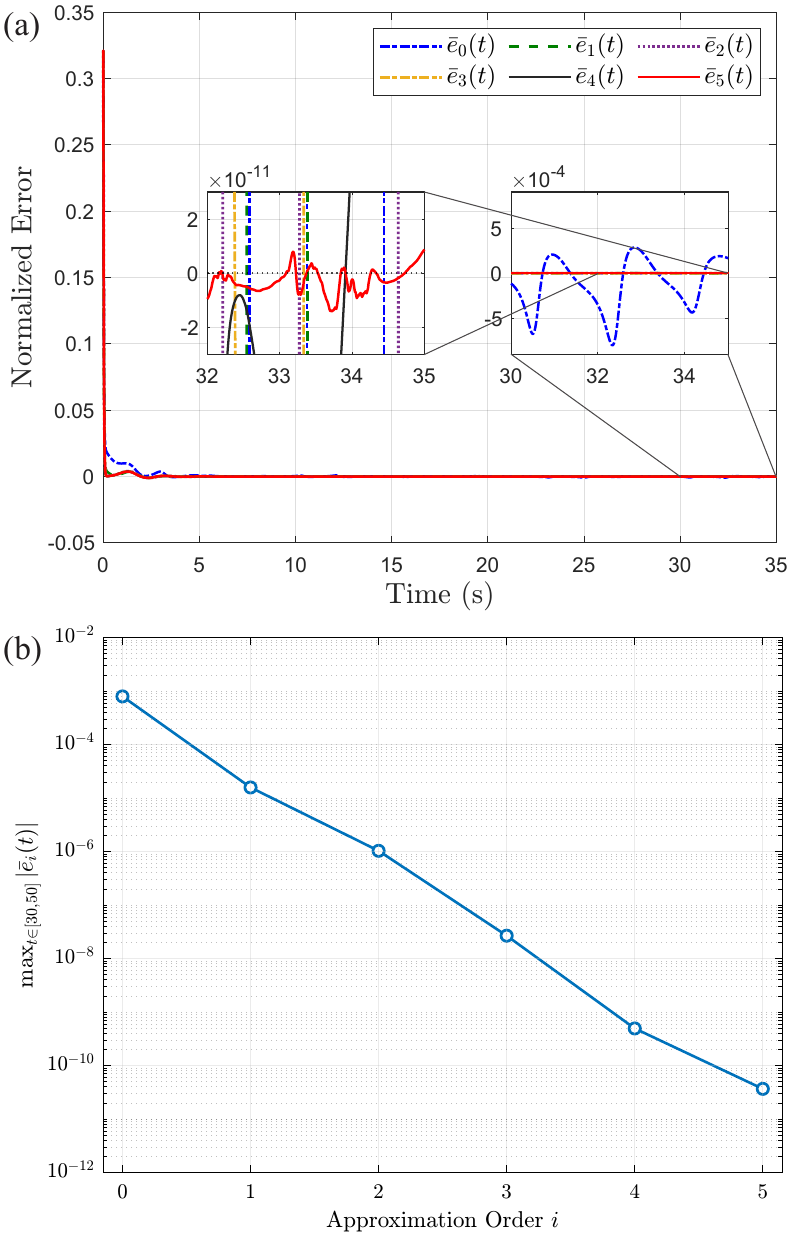}}
\caption{Simulation results of Example~\ref{eg-pu} for singular uncertainty, where $\bar e_i(t),i=0,1,\cdots,5$  denote the normalized errors of the approximate regulators of orders $0,1,\cdots,5$, respectively.
}
\label{fig2}
\end{figure} 
\end{example}

\section{Conclusion and Future Work}
This paper has investigated the fundamental limitation and feasible design routes for robust output regulation of uncertain LTV systems.
By the proposed trajectory-matching system immersion, we have shown that the uncertainties, specifically within the high-frequency gain $b(t,\mu)$ and zero-dynamics $A_\eta(t,\mu)$, can generate infinite-dimensional error-zeroing mappings. 
This elucidates the general failure of finite-dimensional internal models for uncertain LTV systems: 
under the standing assumptions and whenever the corresponding stabilizing realization is feasible, exact robust regulation via a linear finite-dimensional internal-model-based regulator is achievable if and only if the error-zeroing matrix $R(t,\mu)$ admits a finite linear parameterization, a condition that generally fails.
This capability restriction suggests that, for uncertain LTV systems, robust regulation should be viewed not merely as a direct extension of the LTI internal model principle, but as a problem of balancing finite-dimensional realizability and approximation accuracy.
The proposed controller design provides such a finite-dimensional approximate solution, while recovering exact regulation for certain specified uncertainties. 

Future work may investigate more dimension-efficient approximation schemes that retain strongly observable directions, or alternative realization methods with weaker dependence on the observability of the internal model.
When exact regulation is indispensable, infinite-dimensional realizations or nonlinear regulator mechanisms beyond finite linear parameterization may provide possible directions.

\renewcommand{\thesection}{Appendix A}
\section{Discussion on Internal Models}\label{app-im}

\setcounter{lemma}{0}
\setcounter{proposition}{0}
\setcounter{corollary}{0}
\setcounter{remark}{0}
\setcounter{equation}{0}
\setcounter{definition}{0}

\renewcommand{\thedefinition}{A.\arabic{definition}}
\renewcommand{\thelemma}{A.\arabic{lemma}}
\renewcommand{\theproposition}{A.\arabic{proposition}}
\renewcommand{\thecorollary}{A.\arabic{corollary}}
\renewcommand{\theremark}{A.\arabic{remark}}
\renewcommand{\theequation}{A.\arabic{equation}}

A minimum IM means it possesses the minimum dimension among all the equivalent IMs.
An IM $\{\Xi(\cdot),\Gamma(\cdot)\}$  can be reduced to a minimum IM via a two-step procedure, beginning with observability decomposition.
% if it is not completely observable.
According to~\cite[Thm. 7]{weiss1968structure}, if $\max_{t>t_0} \{{\rm rank}(W(t_0,t))\} = \tilde \nu <\bar \nu$ for all $t_0$, there exists a diffeomorphic 
transformation $T(\cdot)$ to decompose $\{\Xi(\cdot),\Gamma(\cdot)\}$ as $\tilde \Xi(\cdot) = [\tilde \Xi_{11}(\cdot),\mathbf{0};\tilde \Xi_{21}(\cdot) , \tilde \Xi_{22}(\cdot)]$, $\tilde \Gamma(\cdot)= [\tilde \Gamma_{1}(\cdot) ,\mathbf{0} ]$,
with $\{\tilde \Xi_{11}(\cdot),\tilde \Gamma_1(\cdot)\}$ completely observable.
If $T(\cdot)$ is a Lyapunov transformation, $\tilde \Xi_{11}\in C^\infty(\mathbb{R}, \mathbb{R}^{\tilde \nu\times \tilde \nu})$ is still marginally stable.
The second step removes the redundancy in the range of initial value $\hat \Sigma_1(t_0,\mu)$ after transformation by projection.
Suppose there exists $t_0\in\mathbb{R}$ and nonzero $\tilde \lambda\in \mathbb{R}^{\tilde \nu}$, such that  $\tilde \lambda^T \hat \Sigma_1(t_0,\mu) = 0$ for all  $\mu\in \mathcal{P}$.
Then we can find $V(t_0)\in \mathbb{R}^{\tilde \nu \times \hat \nu}$ of full column rank such that for any $\mu\in \mathcal{P}$, there exists $\hat \Sigma_2(t_0,\mu) \in \mathbb{R}^{\hat \nu\times \rho} $ satisfying $\hat \Sigma_1(t_0,\mu) =V(t_0) \hat \Sigma_2(t_0,\mu) $.
Moreover, to make $\hat \nu$ as small as possible, we should ensure for any nonzero $\hat \lambda\in \mathbb{R}^{\hat \nu}$, there exists $\mu \in \mathcal{P}$ such that $\hat \lambda^T \hat \Sigma_2(t_0,\mu)\neq 0$.
Denote $V(t) = \Phi_{\bar \Xi_{11} }(t,t_0) V(t_0)$, 
and then $\{\mathbf{0},\tilde\Gamma_1(t)V(t)\}$ is a minimum IM of order $\hat \nu$.
%It is marginally stable and still completely observable, 

\begin{remark}\label{rm:cim}
  The above analysis links the observability (See Supplementary~S1) of IM to its dimension, which is natural by noting the unforced structure~\eqref{eq-isi-im}.
  A minimum IM means there is no redundancy in the initial value $\hat \Sigma(t_0,\mu)$, which can then always be observed from the output of $\{\Xi(\cdot),\Gamma(\cdot)\}$, thus  having the minimum dimension.
  This extends the results in~\cite{cox2016isolating}, which reduce the dimension of IMs by isolating the invisible parts.  However,  since invisibility is stronger than unobservability, this method seems more conservative.
  A regular IM strengthens the requirement on observability to enable the regulator design in Proposition~\ref{pr-cr}, which is named according to~\cite[Def. 3.1]{zhang2009tac}, where system immersion is classified into different types based on the observability of the  immersion system.
\end{remark}

Besides reducing the dimension of an IM, it may also be increased.
Proposition~\ref{pr-er}, whose proof is in Supplementary~S5, states the condition for equivalent IMs, which is indeed immersing $\{\Xi(\cdot),\Gamma(\cdot)\}$ into $\{\bar \Xi(\cdot),\bar \Gamma(\cdot)\}$. 

\begin{proposition}\label{pr-er}
  Two IMs $\{\Xi(\cdot),\Gamma(\cdot)\}$ and $\{\bar \Xi(\cdot),\bar \Gamma(\cdot)\}$ are equivalent, if there exists $L(t)\in C^\infty (\mathbb{R}, \mathbb{R}^{\tilde \nu \times \bar \nu})$ 
%with constant rank $q$, 
such that 
\begin{equation}
  \begin{aligned}
\dot L(t)+ L(t) \Xi(t) &= \bar \Xi(t)L(t)\\
\Gamma(t)&= \bar \Gamma(t) L(t).
    \end{aligned}  \label{eq-er}
\end{equation}
\end{proposition}

\begin{remark}\label{rm-eqim}
Given an IM $\{\Xi(\cdot),\Gamma(\cdot)\}$, more equivalent IMs can be readily obtained.
Considering any $\bar \Xi\in C^\infty (\mathbb{R}, \mathbb{R}^{\bar \nu \times \bar \nu})$ marginally stable, with $\bar \Gamma(t) = \Gamma(t)\Phi_\Xi(t,\bar t_0) \Phi_{\bar \Xi}(\bar t_0,t)$ for any fixed $\bar t_0\in \mathbb{R}$,
$\{\bar \Xi(\cdot),\bar \Gamma(\cdot)\} $ constitutes an equivalent IM, because the solution to~\eqref{eq-er} is $L(t)$ with $L(t_0) = \Phi_{\bar \Xi}(t_0,\bar t_0)\Phi_\Xi(\bar t_0,t_0)$.
In particular, $\{\mathbf{0}, \Gamma(t)\Phi_\Xi(t,\bar t_0)\}$ is an equivalent IM.
\end{remark}

\renewcommand{\thesection}{Appendix B}
\section{Time-varying Canonical Realization}\label{sec-st}
\setcounter{lemma}{0}
\setcounter{proposition}{0}
\setcounter{corollary}{0}
\setcounter{remark}{0}
\setcounter{equation}{0}
\setcounter{definition}{0}

\renewcommand{\thedefinition}{B.\arabic{definition}}
\renewcommand{\thelemma}{B.\arabic{lemma}}
\renewcommand{\theproposition}{B.\arabic{proposition}}
\renewcommand{\thecorollary}{B.\arabic{corollary}}
\renewcommand{\theremark}{B.\arabic{remark}}
\renewcommand{\theequation}{B.\arabic{equation}}

After obtaining an IM $\{\Xi(\cdot),\Gamma(\cdot)\}$,  $\{F(\cdot),G(\cdot),H(\cdot)\}$ is finally chosen to stabilize the closed-loop system through \textit{canonical realization} and high-gain feedback, which are both extended from the periodic version~\cite[Def. 3.2, Prop. 3.4]{zhang2009tac}.

\begin{definition}[Canonical Realization]\label{def-cr}
  Given $\{S(\cdot),R(\cdot,\cdot)\}$, 
  a smooth triple $\{F_{\rm im}(\cdot),G_{\rm im}(\cdot),H_{\rm im}(\cdot)\}$, whose components and the necessary derivatives are UB, is called a canonical  realization for it, if 
$F_{\rm im}\in C^\infty(\mathbb{R}, \mathbb{R}^{l\times l})$ is UAS,
and $\{F_{\rm im}(\cdot)+G_{\rm im}(\cdot)H_{\rm im}(\cdot),H_{\rm im}(\cdot)\}$ is an IM for $\{S(\cdot),R(\cdot,\cdot)\}$. 
\end{definition}

\begin{proposition}\label{pr-stable}
  With Assumptions~\ref{ass-s}-\ref{ass-m}, suppose the triple  $\{F_{\rm im}(\cdot),G_{\rm im}(\cdot),H_{\rm im}(\cdot)\}$ is a canonical realization. 
  Choose $K\in \mathbb{R}^{1\times (r-1)}$ such that $(A_{\rm b} + B_{\rm b}K)$ is Hurwitz, where the pair $\{A_{\rm b},B_{\rm b}\}$ is in  Brunovsky form of order $(r-1)$.
  Choose the gain parameter $g>1$ and $d_i,i=0,\cdots,r-1$, as the coefficients of an arbitrary Hurwitz polynomial, which define 
  \begin{equation}
    M_g = 
      \begin{bmatrix}
        -g d_{r-1}&1&0&\cdots&0\\
        -g^2 d_{r-2}&0&1&\cdots&0\\
        \vdots&\vdots&\vdots&\ddots&\vdots\\
        -g^r d_{0}&0&0&\cdots&0
      \end{bmatrix}
    , \;
    L_g = 
      \begin{bmatrix}
        g d_{r-1}\\
        g^2 d_{r-2}\\
        \vdots\\
        g^r d_{0}
      \end{bmatrix}. \nonumber
  \end{equation}
  Then there exists a constant $k^* > 0$ such that for all $k> k^*$, there exists a constant $g^*(k)>1$, ensuring that for all $g>g^*$, 
  the controller
\begin{equation}
  \begin{aligned}
       \begin{bmatrix}
  \dot  \xi_1 \\\dot \xi_2
  \end{bmatrix}  =&
   \begin{bmatrix}
  M_g &\mathbf{0} \\ 
  -k {\rm sign}(b) G_{\rm im}[-K\;1] & F_{\rm im}+G_{\rm im}H_{\rm im}
  \end{bmatrix} 
   \begin{bmatrix}
  \xi_1 \\ \xi_2
  \end{bmatrix}  
  + 
   \begin{bmatrix}
  L_g \\\mathbf{0}
  \end{bmatrix} e \\
  u =& \begin{bmatrix}
    -k{\rm sign}(b)[-K\; 1] &H_{\rm im}
  \end{bmatrix}  
   \begin{bmatrix}
  \xi_1 \\ \xi_2
  \end{bmatrix} ,
  \end{aligned}\label{eq-hg}
\end{equation}
where $\xi_1\in\mathbb{R}^r,\xi_2\in\mathbb{R}^{\bar \nu}$ and $\nu = r+\bar \nu$, stabilizes the closed-loop system.
\end{proposition}
\begin{proof}
  The complete proof is provided in Supplementary~S6, with key steps summarized here.
  Adopting the Byrnes-Isidori form, 
  construct the Lyapunov transformation $T(t,\mu)$ on $\mathbf{x}={\rm col}(x,\xi_1,\xi_2)$, and denote
  $\hat {\mathbf{x}} ={\rm col}\{\hat x_1,\hat x_2,\hat x_3,\hat \xi_1,\hat \xi_2\}= T(t,\mu)\mathbf{x}$.
  Select the Lyapunov function as 
  \begin{equation}
    \begin{aligned}
        V(t,\hat {\mathbf{x}},\mu) =&c_1 \hat x_1^T P_K \hat x_1 
  + \frac{1}{2} \hat x_2^2+ c_3 \hat x_3^T P_\eta \hat x_3
  + c_4 \hat \xi_1^T P_M \hat \xi_1
  + c_5 \hat \xi_2^T P_F \hat \xi_2,
    \end{aligned} \label{eq-lf}
\end{equation}
where  $P_K, P_M, P_F(t)$ are symmetric. In particular, $\bar a_1 I \leq P_F(t) \leq \bar a_2 I$ and $\dot{P}_F(t) + P_F(t) F_{\rm im} (t)+ F_{\rm im}^T(t) P_F(t) \leq -\bar a_3I$ for some constants $\bar a_1,\bar a_2,\bar a_3>0$; $A_K^T P_K +P_K A_K =M_0^T P_M + P_M M_0= -I  $, with $A_K = A_b+B_b K$ and $M_0$ as picking $g=1$ in $M_g$.
Denote 
$||F_{\rm im}(t)||\leq \phi_F$, $||H_{\rm im}(t)||\leq \phi_H$,$||G_{\rm im}(t)||\leq \phi_G $, $||{\rm d}(G_{\rm im}(t)/b(t,\mu)) /{\rm d} t|| \leq \phi_{\rm d} $ for constants $\phi_F, \phi_H, \phi_G,\phi_{\rm d}>1$.
Choose 
$c_5>5/\bar a_3$, 
$c_3>2[\phi_0(c_5 \bar a_2 \phi_G) ^2+2]/a_3$, 
$c_1>2c_3 \phi_0 a_2^2/a_3 +2\phi_0(c_5 \bar a_2  \phi_G )^2+4$, 
$c_4 = g^{2(r-1)}$,  
and 
\begin{equation}
  \begin{cases}
      k&>\phi_0 \gamma_1 \\
  g &> \phi_0 (k  +  \gamma_2),
  \end{cases} \label{eq-kg}
\end{equation}
where 
$
      \gamma_1 = c_1+[c_5\bar a_2(\phi_F \phi_G  +\phi_{\rm d})]^2
       + \phi_H^2 +\phi_G\phi_H ,$
$      \gamma_2 =   \phi_H^2+\phi_G\phi_H$.
Then
$
  \dot V(t,\hat {\mathbf{x}},\mu) \leq -\hat X^T \hat Q \hat X, 
  $
where $\hat Q \in \mathbb{R}^{5\times 5}>0$ is constant, and $\hat X=[||\hat x_1||,||\hat x_2||, ||\hat x_3||,||\hat \xi_1||,||\hat \xi_2||]^T $.
\end{proof}

\begin{remark}
  The stabilization construction in Proposition~\ref{pr-stable} depends only on the canonical realization and its boundedness.
  It therefore applies equally to asymptotic and approximate IMs, with the resulting regulation accuracy determined by the corresponding residual as shown in~\eqref{eq-apperr}. 
 In particular, Problem~\ref{prob-1} is solved when using an asymptotic IM.
\end{remark}

Canonical realizations are not unique. Existing LTV constructions may require additional structural conditions~\cite[Prop. 3]{marconi2013internal}.
By strengthening observability, Proposition~\ref{pr-cr} provides a direct construction, adapted from the periodic version~\cite[Prop. 5.1]{zhang2006linear}.

\begin{proposition}\label{pr-cr}
  With Assumption~\ref{ass-s}, suppose $\{\Xi(\cdot),\Gamma(\cdot)\}$ is a regular IM. Set $F_{\rm im}(\cdot)=-\alpha_{\rm cr} I - \Xi^T(\cdot)$ with any constant $\alpha_{\rm cr}>0$. Then $F_{\rm im}(\cdot)$ is UAS, and
\begin{equation}
  \dot L(t) + L(t) \Xi(t) = F_{\rm im}(t) L(t)+\Gamma^T(t)\Gamma(t) \label{prop-select}
\end{equation}
admits a UB solution $L(\cdot)$ that is a Lyapunov transformation. 
Finally, a canonical realization is yielded by setting  $G_{\rm im}(\cdot) = \Gamma^T(t)$ and $H_{\rm im}(\cdot) = \Gamma(\cdot) L^{-1}(\cdot)$.
\end{proposition}

\begin{proof}
The complete proof is provided in Supplementary~S7, and here we only state the solution to $\eqref{prop-select}$ satisfies  $L(t) \geq e^{-\alpha_{\rm cr} \phi_\delta} \hat \phi_\Xi^2  \phi_{\rm uco}  I$ for some constants $\hat \phi_\Xi,\phi_{\rm uco},\phi_\delta >0$. 
In particular, for all $t \in \mathbb{R}$,
$
  \int_{t-\phi_\delta}^{t} (\Gamma(\tau)\Phi_\Xi(\tau,t-\phi_\delta))^T \Gamma(\tau)\Phi_\Xi(\tau,t-\phi_\delta)   \,\mathrm{d}\tau  \geq \phi_{\rm uco} I 
$.
\end{proof}

\renewcommand{\thesection}{Appendix C}
\section{Proofs of the Main Results}
\setcounter{lemma}{0}
\setcounter{proposition}{0}
\setcounter{corollary}{0}
\setcounter{remark}{0}
\setcounter{equation}{0}
\setcounter{definition}{0}

\renewcommand{\thedefinition}{C.\arabic{definition}}
\renewcommand{\thelemma}{C.\arabic{lemma}}
\renewcommand{\theproposition}{C.\arabic{proposition}}
\renewcommand{\thecorollary}{C.\arabic{corollary}}
\renewcommand{\theremark}{C.\arabic{remark}}
\renewcommand{\theequation}{C.\arabic{equation}}

\subsection{Proof of Theorem~\ref{thm-re}} \label{pf-thm-re}

Lemmas~\ref{lm-sol1}-\ref{lm-bi} imply the solvability of the RE \eqref{RE1}-\eqref{RE2} and provide the solutions.
Then Lemma~\ref{lm-ub} indicates it is UB from any feasible initial value under Assumption~\ref{ass-m}. 
Finally, Lemma~\ref{lm-au} indicates asymptotic uniqueness of the solutions to~\eqref{REiff1}, thus completing the proof.

\begin{lemma}[Solution and Solvability]\label{lm-sol1}
    With Assumptions~\ref{ass-s}-\ref{ass-r}, 
     a smooth pair $\{\bar \Pi(\cdot,\cdot),\bar R(\cdot,\cdot)\}$ is a solution to~\eqref{RE1}-\eqref{RE2} if and only if it takes the form of \eqref{eq-pi}-\eqref{eq-R} with $\bar \Pi(t_0,\mu)$ satisfying \eqref{eq-iv} for some $t_0\in \mathbb{R}$. 
     Moreover,~$\{\bar \Pi(\cdot,\cdot),\bar R(\cdot,\cdot)\}$ always exist.
\end{lemma}

\begin{proof}
Necessity.
Calculate the derivatives of $\eqref{RE2}$ and note Assumption~\ref{ass-r}, yielding the first $(r-1)$ derivatives of~$(C(\cdot,\cdot)\bar \Pi(\cdot,\cdot)+Q(\cdot,\cdot))$ satisfying
\begin{equation}
    \mathscr{O}_{A}(t,\mu)\bar \Pi (t,\mu)+ \mathscr{O}_{S}(t,\mu)+\mathscr{P}(t,\mu) =0, \label{eq-de}
\end{equation}
which indicates \eqref{eq-iv}, with the $r$-th derivative as~\eqref{eq-R}.
Substitute~\eqref{eq-R} into \eqref{RE1} and right-multiply both sides by $\Phi_S(t,t_0)$. This transforms the Sylvester  differential equation \eqref{RE1} into the differential equation
\begin{equation}
  \dot {\bar \Psi} (t,t_0,\mu)  = M(t,\mu)\bar \Psi(t,t_0,\mu) +N(t,\mu)\Phi_S(t,t_0), \label{eq-psi}
\end{equation} 
where $\bar\Psi (t,t_0,\mu)=\bar \Pi(t,\mu) \Phi_S(t,t_0) $, thus yielding~\eqref{eq-pi}.

Sufficiency. 
Since \eqref{eq-psi} always admits a smooth solution~\eqref{eq-pi}, and $\bar R(\cdot,\cdot)$ can be calculated using \eqref{eq-R}, it remains to prove such $\bar \Pi(\cdot,\cdot)$ satisfies \eqref{RE2}.
That $\bar \Pi(t_0,\mu)$ satisfying~\eqref{eq-iv} means the first $(r-1)$ derivatives of $(C(\cdot,\cdot)\bar \Pi(\cdot,\cdot)+Q(\cdot,\cdot))$ at $t_0$ are zero.
Because \eqref{eq-R} enforces the $r$-th derivative to remain zero,~\eqref{eq-de}, and hence~\eqref{RE2}, hold for all $t$. 

According to \cite[Prop. 4.5]{pinzoni1989stabilization}, ${\rm rank}(\mathscr{O}_A (\cdot,\cdot))\equiv r$, so \eqref{eq-iv} is solvable for any $\{P(\cdot,\cdot),Q(\cdot,\cdot)\}$, completing the proof.
\end{proof}

\begin{lemma}\label{lm-bi}
    With Assumptions~\ref{ass-s}-\ref{ass-r},  adopting the Byrnes-Isidori form, the solution is~\eqref{eq-pi-BIu} and 
    \begin{equation}
        \begin{aligned}
        &\bar \Pi_{\rm BI, l}(t,\mu) =\Phi_{\eta}(t,t_0,\mu)\bar \Pi_{\rm BI,l}(t_0,\mu) \Phi_S(t_0,t)
        + \int_{t_0}^{t} \Phi_{\eta}(t,\tau,\mu) ( P_{\rm BI, l}(\tau,\mu) - \beta(\tau,\mu) Q(\tau,\mu))\Phi_S(\tau,t)  \,\mathrm{d}\tau ,
        \end{aligned} \label{eq-pil}
    \end{equation}
    where $\bar \Pi_{\rm BI,l}(t_0,\mu)$ is any constant matrix in $\mathbb{R}^{(n-r)\times \rho}$.
\end{lemma}
\begin{proof}
  The complete proof is in Supplementary~S8, and here we only point out that \eqref{eq-pil} satisfies 
\begin{equation}
    \dot {\bar \Pi}_{\rm BI,l}+\bar \Pi_{\rm BI,l}S = A_\eta \bar \Pi_{\rm BI,l}-\beta Q +P_{\rm BI,l}. \label{eq-se-sim}
\end{equation}
\end{proof}

\begin{lemma}[Uniform Boundedness]\label{lm-ub}
    With Assumptions~\ref{ass-s}-\ref{ass-m}, the solution to RE \eqref{RE1}-\eqref{RE2} is always UB.
\end{lemma}

\begin{proof}
    Take Byrnes-Isidori form. 
    Since $A_\eta(\cdot,\cdot)$ is UAS, $\bar \Pi(\cdot,\cdot)$ and then $\bar R(\cdot,\cdot)$ in Lemma~\ref{lm-bi} are UB.
\end{proof}

Now consider uniqueness. Though there may exist many solutions to~\eqref{REiff1}, they are unique in the sense of asymptotic behavior. 
This is shown in Lemma~\ref{lm-au}, and we provide Lemma~\ref{lm-Barbalat} first with its proof in Supplementary~S9.

\begin{lemma}\label{lm-Barbalat}
With assumptions~\ref{ass-r}-\ref{ass-m}, consider a SISO LTV plant $\{A(t),B(t),C(t)\}$.
If  the input $u(t)$ and $\dot u(t)$ are UB, and at least one of the following conditions holds:
\begin{enumerate}[(i)]
  \item the first $(r-1)$ derivatives of the output $y(t)$ are UB;
  \item $A(\cdot)$ is UAS,
\end{enumerate}
then  $y \to 0(t \to \infty)$ indicates $x,u\to 0$.
\end{lemma}

\begin{lemma}[Asymptotic Uniqueness]\label{lm-au}
With Assumptions~\ref{ass-s}-\ref{ass-m}, 
if at least one of the following conditions holds:
\begin{enumerate}[(i)]
  \item the first $k$ derivatives of $R(\cdot,\cdot)$ are UB with $k=\max\{1,r-2\}$;
  \item $\dot R(\cdot,\cdot)$ is UB, and $A(\cdot,\cdot)$ is UAS,
\end{enumerate}
then all the UB solutions to~\eqref{REiff1} are asymptotically unique, meaning their difference decays to zero when $t\to \infty$.
In particular, they all converge to the solution to \eqref{RE1}-\eqref{RE2}, namely $\{\bar \Pi(\cdot,\cdot),\bar R(\cdot,\cdot)\}$.
\end{lemma}

\begin{proof}
  We prove the case for condition (i), and the other case is analogous.
Lemma~\ref{lm-sol1} indicates the solutions to \eqref{RE1}-\eqref{RE2} always exist, and Lemma~\ref{lm-ub} shows all of them and then their derivatives are UB.
Pick one  denoted as $\{\bar \Pi_1(\cdot,\cdot),\bar R_1(\cdot,\cdot)\}$.
Then we show the solutions to~\eqref{REiff1} including other solutions to~\eqref{RE1}-\eqref{RE2} all converge to it.
Suppose~\eqref{REiff1} admits a UB solution $\{\Pi_2(\cdot,\cdot),R_2(\cdot,\cdot)\}$ with the first $k$ derivatives of $R_2(\cdot,\cdot)$ UB, which ensures the first $(k+1)$ derivatives of $\Pi_2(\cdot,\cdot)$ UB.
Denote $\Pi_\delta(\cdot,\cdot) =\bar \Pi_1(\cdot,\cdot) - \Pi_2(\cdot,\cdot)$ and $R_\delta(\cdot,\cdot) =\bar R_1(\cdot,\cdot) - R_2(\cdot,\cdot)$, and
subtraction between~\eqref{RE1}-\eqref{RE2} and~\eqref{REiff1} yields
\begin{equation}
    \begin{aligned}
         \dot {  \Psi}_\delta (t,t_0,\mu) &= A(t,\mu) {  \Psi}_\delta (t,t_0,\mu) +B(t,\mu)\Omega_\delta(t,t_0,\mu) \\
  0 &=  \lim_{t\to\infty} C (t,\mu){  \Psi}_\delta (t,t_0,\mu),
    \end{aligned} \nonumber
\end{equation}
by right-multiplying it by $\Phi_S(t,t_0)$ and denoting ${  \Psi}_\delta(t,t_0,\mu) = \Pi_\delta(t,\mu) \Phi_S(t,t_0)$ and $\Omega_\delta(t,t_0,\mu) = R_\delta(t,\mu) \Phi_S(t,t_0)$.
According to Lemma~\ref{lm-Barbalat}, when $t\to\infty$, $\{\Psi_\delta(\cdot,\cdot,\cdot),\Omega_\delta(\cdot,\cdot,\cdot)\} \to 0$, meaning $\{\Pi_\delta(\cdot,\cdot), R_\delta(\cdot,\cdot)\}\to 0$.
Therefore, $\{\Pi_2(\cdot,\cdot),R_2(\cdot,\cdot)\}$ converges to $\{\bar \Pi_1(\cdot,\cdot), \bar R_1(\cdot,\cdot)\}$. Furthermore, the difference between solutions to \eqref{REiff1} vanishes as well.
\end{proof}

\begin{remark}\label{rm-au}
  If the system~\eqref{sys} is $T$-periodic, then the $T$-periodic solution $\bar \Pi(\cdot,\cdot)$ in \eqref{eq-pi-BI} is unique by setting $t_0$ as~$-\infty$, with proof similar to \cite[Lem. A.1]{zhang2006linear}.
Besides, if $r=n$, then~$\mathscr{O}_{A}(\cdot,\cdot)$ is invertible~\cite[Prop. 4.5]{pinzoni1989stabilization}, which makes the initial value in~\eqref{eq-iv} unique, thus leading to the uniqueness of the solution to the strengthened RE~\eqref{RE1}-\eqref{RE2}. Indeed, $\bar \Pi(\cdot,\cdot)$ can be directly calculated from~\eqref{eq-iv} without bothering with the convolution in~\eqref{eq-pi}.
\end{remark}

\subsection{Proof of Theorem~\ref{thm-flp}} \label{pf-thm-flp}
\begin{proof}
Sufficiency. With the IM~\eqref{eq-rim},
choose 
$ \hat \Sigma(t_0,\mu) =  \left [1,\theta_{R1}(\mu),\cdots,\theta_{R N_R}(\mu)\right ]^T \otimes I_\rho$,
and then $R(t,\mu) = \Gamma(t)\hat \Sigma(t,\mu) +\Delta_{\rm d}(t,\mu)$, making $\{\Xi(\cdot),\Gamma(\cdot)\}$ an asymptotic IM.

  Necessity.
  Denote the UB mapping associated with~$\{\Xi(\cdot),\Gamma(\cdot)\}$ as $\hat\Sigma(t,\mu)$ satisfying \eqref{eq-imdef1} and $ R(t,\mu) = \Gamma(t)\hat \Sigma(t,\mu) + \Delta_R(t,\mu)$, where $\Delta_R(t,\mu)\to \mathbf{0}$.
  Fix $t_0$ and define $\mathcal{V}={\rm span}\{\hat \Sigma(t_0,\mu)-\hat \Sigma(t_0,\mathbf{0}): \mu \in \mathcal{P}\}$.
  Since $\mathcal{V}\subseteq \mathbb{R}^{\bar \nu \times \rho}$, let $m_{\mathcal{V}} = {\rm dim}\mathcal{V}<\infty$, and choose $\mu_k\in \mathcal{P}$, $k=1,\cdots, m_{\mathcal{V}}$ such that $Z^0_k = \hat \Sigma(t_0,\mu_k)-\hat \Sigma(t_0,\mathbf{0})$ form a basis of $\mathcal{V}$.
  Then $\hat \Sigma(t_0,\mu)-\hat \Sigma(t_0,\mathbf{0})=\sum_{k=1}^{m_\mathcal{V}}\theta_{Rk}(\mu)Z_k^0 $, where the scalar coordinate functions $\theta_{Rk}(\mu)$ are smooth.
  Hence, $\hat \Sigma(t,\mu)=\hat \Sigma(t,\mathbf{0})+\sum_{k=1}^{m_\mathcal{V}}\theta_{Rk}(\mu)Z_k $,
  where $Z_k(t)= \Phi_{\Xi}(t,t_0)Z^0_k \Phi_S(t_0,t)=\hat \Sigma(t,\mu_k)-\hat \Sigma(t,\mathbf{0})$ is UB.
  Setting $R_k(t)=\Gamma(t)Z_k(t)$ and $R_0(t)=R(t,\mathbf{0})$, we obtain
  $ R(t, \mu) =R_0(t)+ \sum_{k=1}^{m_{\mathcal{V}}} \theta_{Rk}(\mu) R_k(t)+ \Delta_R(t,\mu)- \Delta_R(t,\mathbf{0})$. 
  The last term decays to zero. 
  Absorbing $\theta_{Rk}(\mu) R_k(t)$ that decays into the residual, and reordering the remaining terms yields~\eqref{eq-flp}.
  For the periodic case, the asymptotic IM is exact, so $\Delta_R(\cdot,\cdot)\equiv0$ and $R_k(t)=R(t,\mu_k)-R(t,\mathbf{0})$, yielding periodic $R_k(\cdot)$ and $\Delta_d(\cdot,\cdot)\equiv0$.
\end{proof}

\subsection{Proof of Corollary~\ref{cr-flp}} \label{pf-cr-flp}
\begin{proof}
Denote the embedded IM as~$\{\Xi(\cdot),\Gamma(\cdot)\}$, and the immersion mapping as $L_{si}(t)$.

Sufficiency.  
Since the embedded IM is asymptotic,
there exists a UB mapping $\hat \Sigma(t,\mu)$ satisfying~\eqref{eq-imdef1}-\eqref{eq-as-im}.
Denote $\bar \Sigma(t,\mu)=L_{\rm si}(t)\hat \Sigma(t,\mu)$, satisfying~\eqref{RE3} and $H(\cdot)\bar\Sigma(\cdot)=\Gamma(\cdot)\hat \Sigma(\cdot)$.
Since $R(t,\mu)$ converges to $\bar R(t,\mu)$,
$\bar R(\cdot,\cdot) = H(\cdot)\bar \Sigma(\cdot) +\bar \Delta_R(\cdot,\cdot)$ with $\bar \Delta_R(\cdot,\cdot)\to 0$ $(t\to \infty)$.
The Lyapunov transformation $\tilde{x} = x-\bar \Pi(t,\mu) w, \tilde{\xi} = \xi-\bar \Sigma(t,\mu) w$ yields
  \begin{equation}
    \begin{bmatrix}
      \dot{\tilde{x}}\\
      \dot{\tilde{\xi}}
    \end{bmatrix} =
    A_{\rm cl}(t,\mu)
    \begin{bmatrix}
      \tilde{x}\\
      \tilde{\xi}
    \end{bmatrix} 
    - \begin{bmatrix}
      B(t,\mu) \bar \Delta_R(t,\mu)\\
     0
    \end{bmatrix}  w .
    \nonumber
  \end{equation}
  Because $e = C(t,\mu)\tilde{x}$, we obtain  that when $t\to \infty$,
  \begin{equation}
    \begin{aligned}
          e(t) \to &
    -\begin{bmatrix} C(t,\mu) &\mathbf{0}\end{bmatrix} 
    %\int_{t_0}^{t}  
    \int_{-\infty}^{t}  
    \Phi_{\rm cl}(t,\tau,\mu)  
    \begin{bmatrix}
      B(\tau,\mu) \bar \Delta_R(\tau,\mu)\\
     \mathbf{0}
    \end{bmatrix} 
    %\Phi_S(\tau,t_0) w(t_0)
    w(\tau)
     \,\mathrm{d}\tau .
    \end{aligned} \label{eq-apperr}
    \end{equation}
  Since $\bar \Delta_R(\cdot,\cdot) \to 0$ and $A_{\rm cl}(\cdot,\cdot)$ is UAS,  $e\to 0$.

  Necessity. 
  Since $R(\cdot,\cdot)=H(\cdot)\Sigma(\cdot,\cdot)$ by Proposition~\ref{pr-re}, $[R(\cdot,\cdot)-H(\cdot)\bar\Sigma(\cdot,\cdot)] \to 0$ when $t\to \infty$. 
    Since $\bar \Sigma(\cdot,\cdot)$ is induced by the embedded IM, there exists a UB mapping satisfying~\eqref{eq-imdef1} such that $\bar \Sigma(t,\mu)=L_{\rm si}(t)\hat \Sigma(t,\mu)$ with $\Gamma(\cdot) = H(\cdot)L_{\rm si}(\cdot)$.
    Hence, $[R(\cdot,\cdot)-\Gamma(\cdot)\hat\Sigma(\cdot,\cdot)] \to 0$, showing $\{\Xi(\cdot),\Gamma(\cdot)\}$ is an asymptotic IM.

According to Theorem~\ref{thm-flp}, the proof is complete.
\end{proof}

\subsection{Proof of Lemma~\ref{lm-infb}}\label{pf-lm-infb}
\begin{proof}
  We omit the subscript ``${\rm BI}$'' throughout this proof.
  Theorem~\ref{thm-re} indicates
  \begin{equation}
   \bar R(t,\bar \mu ) = \frac{\bar R_0(t)}{1+\bar \mu \bar E_b (t)}. \label{eq-r-infb}
  \end{equation}
  Suppose, to the contrary, that a linear finite-dimensional IM exists. By Theorem~\ref{thm-flp}, the non-decaying components of $  \{\bar R(\cdot,\bar \mu ):\bar \mu \in\mathcal I_b\}$ 
  belong to a finite-dimensional function space. Hence, there exist distinct parameters $\bar \mu_1,\ldots,\bar \mu_m\in\mathcal I_b$  and constants $d_1,\ldots,d_m$, not all zero, such that $
\lim_{t\to\infty} \sum_{j=1}^{m}d_j\bar R(t,\bar \mu_j) =0$.
Right-multiply~\eqref{eq-r-infb} by $v_b$ to yield
\begin{equation}
 \lim_{t\to\infty} \bar R_0(t)v_b
\sum_{j=1}^{m}
\frac{d_j}{1+\bar \mu_j \bar E_b (t)} =0. \label{eq-lim-infb}
\end{equation}
Fix any $\bar x \in\bar {\mathcal I}$. Since $\bar {\mathcal I}\subset \bar E_b (\mathcal J_\ell)$ for every $\ell$, there exists $t_\ell\in\mathcal J_\ell$ such that $\bar E_b (t_\ell)=\bar x $.
Moreover, $t_\ell\geq a_\ell\to+\infty$. 
Considering $|\bar R_0(t_\ell)v_b|\geq\phi_R>0$, evaluate~\eqref{eq-lim-infb} along this sequence to yield
$  \sum_{j=1}^{m}
d_j/(1+\bar \mu_j \bar x )\equiv 0$ for 
$\bar x \in\bar {\mathcal I}$. 
By Assumption~\ref{ass-r}, the left-hand side is a rational function of $\bar x $  vanishing on the nondegenerate interval $\bar {\mathcal I}$, so it must vanish identically. By the uniqueness of the partial-fraction decomposition and the distinctness of $\bar \mu_1,\ldots,\bar \mu_m$, it follows that $d_1=\cdots=d_m=0$, which contradicts their construction.
Thus, every finite collection of distinct members of $\{\bar R(\cdot,\bar \mu ):\bar \mu \in\mathcal I_b\}$ is linearly independent modulo functions that decay to zero. 
And Theorem~\ref{thm-re} indicates that every UB solution $R(\cdot,\bar \mu )$ of the original RE differs from $\bar R(\cdot,\bar \mu )$ only by a decaying function. Therefore, the finite linear parameterization in Theorem~\ref{thm-flp} is impossible for $R(\cdot,\cdot)$.
\end{proof}

\subsection{Proof of Lemma~\ref{lm-inf}}\label{pf-lm-inf}
\begin{proof}
  We omit the subscript ``${\rm BI}$'' throughout this proof.
Define $Z(t,\bar \mu )=T_\eta^{-1}(t)\bar \Pi_{\rm l}(t,\bar \mu )T_S(t)$, and consider that the Floquet factors satisfy $\dot T_\eta(t)=A_{\eta0}(t)T_\eta(t)-T_\eta(t)\lambda_\eta$ and $\dot T_S(t)=S(t)T_S(t)-T_S(t)S_0$. Then~\eqref{eq-se-sim} gives
\begin{equation}
  \dot Z(t,\bar \mu )+Z(t,\bar \mu )S_0=(\lambda_\eta+\bar \mu )Z(t,\bar \mu )+U(t). \label{eq-z}
\end{equation}

By Assumption~\ref{ass-s},  $S_0$ is diagonalizable over $\mathbb C$, and all its eigenvalues have zero
real parts. Let $v_1,\ldots,v_\rho$ be an eigenvector basis of $S_0$. For every $k$ such that $U_k\neq0$, there exists at least one $v_i$ satisfying $U_kv_i\neq0$. Otherwise, $U_k$ would vanish on a basis of $\mathbb C^\rho$, implying $U_k=0$. Since infinitely many $U_k$ are nonzero while there are only finitely many eigenvectors, there exists an eigenvector $v_\eta\in\mathbb C^\rho\setminus{\mathbf{0}}$ such that $U_kv_\eta\neq0$ for infinitely many $k$. Denote its eigenvalue by $\lambda_S$, namely $S_0v_\eta=\lambda_Sv_\eta$.

Define $z(t,\bar \mu )=Z(t,\bar \mu )v_\eta, u(t)=U(t)v_\eta$, yielding $u(t)=\sum_{k\in\mathbb Z}u_ke^{\mathrm jk\omega t}$, with $u_k=U_kv_\eta$. 
Right-multiplying~\eqref{eq-z} by $v_\eta$ yields
$\dot z(t,\bar \mu )=(\lambda_\eta+\bar \mu -\lambda_S)z(t,\bar \mu )+u(t)$,
whose unique periodic solution  has Fourier coefficients
$  z_k(\bar \mu )=u_k/(\mathrm jk\omega+\lambda_S-\lambda_\eta-\bar \mu )$. 
The denominators do not vanish, because Assumption~\ref{ass-m} implies that $(\lambda_\eta+\bar \mu )$ has negative real part, whereas $(\mathrm jk\omega +\lambda_S)$ have zero real parts.

Take any distinct parameters $\bar \mu_1,\ldots,\bar \mu_m\in\mathcal I_\eta$ and suppose that there exist $d_1,\ldots,d_m\in\mathbb C$ such that $\sum_{i=1}^{m}d_iz(t,\bar \mu_i )\equiv0$. Then for every $k$ satisfying $u_k\neq0$,  $\sum_{i=1}^{m}d_i/(\mathrm jk\omega+\lambda_S-\lambda_\eta-\bar \mu_i )=0$.
Therefore, the rational function $g(s)=\sum_{i=1}^{m}d_i/(s-\bar \mu_i )$ vanishes at infinitely many distinct points $s=\mathrm jk\omega+\lambda_S-\lambda_\eta$, and hence $g\equiv0$ (similar to Supplementary~S3). 
Following~\ref{pf-lm-infb},
$d_i=0$ for every $i$. 
Thus, every finite collection of distinct members of $\{z(\cdot,\bar \mu ):\bar \mu \in\mathcal I_\eta\}$ is linearly independent. 
Consequently, $\{Z(\cdot,\bar \mu ):\bar \mu \in\mathcal I_\eta\}$ and $\{\Pi_{\rm l}(\cdot,\bar \mu ):\bar \mu \in\mathcal I_\eta\}$ both span infinite-dimensional function spaces.

It remains to show that this infinite-dimensional variation is visible in $R(\cdot,\cdot)$. Since only $A_\eta(t,\mu)$ is uncertain, $\Pi_{\rm u}(t)$ is independent of $\bar \mu $, and then 
\begin{equation}
  R(t,\bar \mu )= R_c(t)-b^{-1}(t)\alpha_\eta(t) \Pi_{\rm l}(t,\bar \mu ), \label{eq-R-c}
\end{equation}
where $R_c(t)$ is independent of $\bar \mu $.
Suppose, to the contrary, that $\{ R(\cdot,\bar \mu ):\bar \mu \in\mathcal I_\eta\}$ spans a finite-dimensional function space. Fix any $\bar \mu_0 \in\mathcal I_\eta$. Then the differences $( R(\cdot,\bar \mu )- R(\cdot,\bar \mu_0 ))$ also span a finite-dimensional space. 
The assumption on $\alpha_\eta(\cdot)$ implies that the set $\{t\in[0,T]:\alpha_\eta(t)\neq0\}$ is dense in $[0,T]$, which means if $b^{-1}(t)\alpha_\eta(t)f(t)\equiv0$, then $f(t)=0$ wherever $\alpha_\eta(t)\neq0$, and continuity yields $f\equiv0$.
So $\{\Pi_{\rm l}(\cdot,\bar \mu )-\Pi_{\rm l}(\cdot,\bar \mu_0 ):\bar \mu \in\mathcal I_\eta\}$, and then $\{\Pi_{\rm l}(\cdot,\bar \mu ):\bar \mu \in\mathcal I_\eta\}$, spans a finite-dimensional space, leading to the contradiction.
Thus, $\{R(\cdot,\bar \mu ):\bar \mu \in\mathcal I_\eta\}$ is infinite-dimensional. Considering the system is $T$-periodic and according to Corollary~\ref{cr-flp}, $ R(\cdot,\cdot)$ cannot satisfy~\eqref{eq-flp}, so Problem~\ref{prob-1} cannot be solved by a linear finite-dimensional IM-based regulator.
\end{proof}

\subsection{Proof of Lemma~\ref{lm-inf2}}\label{pf-lm-inf2}
\begin{proof} 
  We only sketch the argument here; the complete proof is
given in Supplementary~S10.
  After the same Floquet change of coordinates as~\ref{pf-lm-inf}, a scalar component satisfies
 \begin{equation} 
  \dot {\hat z}(t,\bar \mu ) =\big(\lambda_\eta+\bar \mu E_\eta(t)-\hat\lambda_S\big)\hat z(t,\bar \mu )+\hat u(t),\nonumber
\end{equation}
Suppose  $\{R(\cdot,\bar \mu ):\bar \mu \in\mathcal I_\eta\}$ spans a finite-dimensional function space. 
Then, for pairwise distinct $\bar \mu_0 ,\ldots,\bar \mu_m \in\mathcal I_\eta$, there would exist constants $d_i$, not all zero, such that $\sum_{i=0}^{m}d_i=0$ and 
$\sum_i d_i \hat z(t,\bar\mu_i)=0$ in a neighborhood of $t^*$.
Define
$F_q(t)=\sum_{i=0}^{m} d_i(\bar \mu_i )^q \hat z(t,\bar \mu_i )$, and
$ c_q=\sum_{i=0}^{m} d_i(\bar \mu_i )^q$.
Differentiating gives
 \begin{equation} 
  \dot F_q(t) =(\lambda_\eta-\hat\lambda_S)F_q(t)+E_\eta(t)F_{q+1}(t)+\hat u(t)c_q. \nonumber
\end{equation}
Since $F_0(t)\equiv0$, $c_0=0$, $E_\eta(t_*)=0$, $\dot E_\eta(t_*)\neq0$, and $\hat u(t^*)\neq0$, induction yields $c_q=0$ for $q=0,1,\ldots,m$. The Vandermonde matrix associated with the distinct $\bar\mu_i$ is nonsingular, so $d_i=0$ for every $i$, contradicting their construction.
Thus, the family is infinite-dimensional, and Theorem~\ref{thm-flp} precludes a linear finite-dimensional IM.
\end{proof}

\subsection{Proof of Theorem~\ref{thm-rbpq}}\label{pf-thm-rbpq}

\begin{proof}
  Throughout the proof, the subscripts ``${Pi}$'' and ``${Qi}$'' mean the matrices are calculated by replacing $\{P(\cdot,\cdot),Q(\cdot,\cdot)\}$ with  $\{E_{Pi}(t),\mathbf{0}\}$ and $\{\mathbf{0},E_{Qi}(t)\}$, respectively.
  The uncertainty $\mu$ only influences $\mathscr{O}_S(t,\mu) ,\bar {\mathscr{O}}_{S}(t,\mu),\mathscr{P}(t,\mu) ,\bar {\mathscr{P}}(t,\mu),N(t,\mu),\bar N(t,\mu)$, which are all linear in $\{P(\cdot,\cdot),Q(\cdot,\cdot)\}$.
  Considering \eqref{eq-iv}, the initial value $\bar \Pi(t_0,\mu)$ is also linear in $\mathscr{O}_S(t_0,\mu)$ and $\mathscr{P}(t_0,\mu)$. 
  Finally, \eqref{eq-R} yields
  \begin{equation}
      \bar R(t,\mu) = \bar R_0(t) 
      + \sum_{i=1}^{N_Q}\theta_{Q i}(\mu)\bar   R_{Qi}(t)
       + \sum_{i=1}^{N_P}\theta_{P i}(\mu)\bar  R_{Pi}(t), \nonumber
  \end{equation}
  where $\bar R_{Qi}(\cdot) $ and $\bar R_{Pi}(\cdot)$ correspond to the initial value $\bar \Pi_{Qi}(t_0)$ and $\bar \Pi_{Pi}(t_0)$, respectively.
  Theorem~\ref{thm-re} ensures that they are all UB.
  Therefore, Theorem~\ref{thm-flp} yields an IM as~\eqref{eq-sim-pq}.
\end{proof}

\subsection{Proof of Proposition~\ref{pr-app}}\label{pf-pr-app}
\begin{proof} 
  Uniform analyticity with respect to \(\mu\) ensures
\begin{equation}
  \begin{aligned}
      \sup_{t \in \mathbb{R}} \left| D_\mu^m (b^{-1})(t, \zeta) \right| &\le \frac{1}{\tilde {\phi}_b} m! \, C_b^{|m|}, \quad \forall \zeta \in \mathcal{P},\\
    \sup_{t \ge t_0} \left\| D_\mu^m \Phi_\eta(t, t_0, \zeta) \right\| &\le \check \phi_\eta e^{-\bar \phi_\eta(t-t_0)} m! \, C_\eta^{|m|}, \, \forall \zeta \in \mathcal{P},
  \end{aligned}
  \label{eq-cauchy_estimate}
\end{equation}
with details in Supplementary~S11.
  The Taylor expansion of $b^{-1}(t,\mu)$ around $\mu=\mathbf{0}$ with the Lagrange remainder yields:
\begin{equation}
    b^{-1}(t,\mu) = \sum_{0 \le |m| \le k_b} \frac{D_\mu^m (b^{-1})(t,\mathbf{0})}{m!} \mu^m + \Delta_b^{k_b}(t,\mu), \nonumber
\end{equation}
where 
$
    \Delta_b^{k_b}(t,\mu) = \sum_{|m| = k_b+1} D_\mu^m (b^{-1})(t,\tilde \zeta) \mu^m/m!$, 
for some $\tilde \zeta \in \mathcal{P}$. Substituting~\eqref{eq-cauchy_estimate} yields:
\begin{equation}
    |\Delta_b^{k_b}(t,\mu)| \le  \frac{1}{\tilde {\phi}_b} (C_b \phi_\mu)^{k_b+1} \binom{N+k_b}{k_b+1}, \nonumber
\end{equation}
where the combinatorial number satisfies $\binom{N+k_b}{k_b+1} \le M_b (\bar{C}_b / C_b)^{k_b+1}\tilde {\phi}_b$ for all $k_b \ge 0$. 
Therefore, $\Delta_b^{k_b}(t,\mu)$ satisfies~\eqref{eq-bds}.
The proof for $\Phi_\eta(\cdot,\cdot,\cdot)$ is similar and is omitted.
\end{proof}

\subsection{Proof of Theorem~\ref{thm-rbapp}} \label{pf-thm-rbapp}
Without loss of generality, always choose $\phi_{\rm uco}\le 1$ for simplicity.
We first provide Proposition~\ref{pr-bd} with its proof in Supplementary~S12.

\begin{proposition}\label{pr-bd}
  Considering $b_i^{-1}(t)$ and $\bar  \Pi_i(t)$ designed in Proposition~\ref{pr-app} and Remark~\ref{rm-cal}, suppose $ \dot b(t,\mu) $  is a uniformly real-analytic function of $\mu \in \mathcal{U}$,
  and choose $\tilde{C}_\eta $ sufficiently large, independently of $k_\eta$.
  Then  there exist constants $\bar {\phi}_b,\phi_\Pi>0$ such that $\sup_t ||b_{i}^{-1}(t)|| \le \bar {\phi}_b$, 
  $\sup_t ||\mathrm{d} b_{i}^{-1}(t)/\mathrm{d} t|| \le \bar {\phi}_b$ 
  for $i=0,1,\cdots,\bar N_b$;
  and $\sup_t ||\bar\Pi_{i}(t)|| \le \phi_\Pi$, $\sup_t ||\dot {\bar\Pi}_{i}(t)|| \le \phi_\Pi$ for $i=0,1,\cdots,\bar N_\eta$.
\end{proposition}

  The stabilization in Proposition~\ref{pr-stable} still works for any approximation order $\{k_b,k_\eta\}$, so  $A_{\rm cl}(t,\mu)$ is UAS and 
  $||\Phi_{\rm cl} (t,s,\mu)||\leq \phi_{\rm cl} e^{-\alpha_{\rm cl} (t-s)},\; \forall t\geq s$
  for some constants $\phi_{\rm cl}(k_b,k_\eta), \alpha_{\rm cl}(k_b,k_\eta)>0 $. 
  %which may depend on the approximation order.
  Similar to~\eqref{eq-apperr},   
\begin{equation}
 \limsup_{t\to\infty} |e(t)| \leq \phi_0 \phi_w \frac{\phi_{\rm cl}(k_b,k_\eta)} {\alpha_{\rm cl}(k_b,k_\eta)}\sup_{t,\mu\in\mathcal{P}} \left|\left| \Delta_R^{k_b,k_\eta}(t,\mu)\right |\right| .
  \label{eq-et}
\end{equation}
It remains to estimate $ \alpha_{\rm cl}(k_b,k_\eta) $ and $\phi_{\rm cl}(k_b,k_\eta)$.

Recall the  Lyapunov function~\eqref{eq-lf}, 
and pick the undetermined coefficients as twice of their lower bounds, namely 
$c_5 = 10/\bar a_3$, 
$c_3 = 4[\phi_0(c_5 \bar a_2  \phi_G) ^2+2]/a_3$, 
$c_1 = 4[c_3\phi_0 a_2 ^2/a_3 +\phi_0(c_5 \bar a_2  \phi_G )^2+2]$. 
Considering $ \alpha_{\rm cl}(k_b,k_\eta) $, we have shown $\dot{V}(t,{\mathbf{x}},\mu) \le -\gamma V(t,{\mathbf{x}},\mu)$ for some constant $\gamma>0$. Note the diagonal structure of $\hat P(t,\mu)$ and $\hat{Q}$, and obtain
\begin{equation}
  \gamma = \inf_{t,\mu\in \mathcal{P}}  \min_{x \neq 0} \frac{\hat{X}^T \hat Q \hat{X}}{\hat{\mathbf{x}}^T \hat P(t,\mu) \hat{\mathbf{x}}} \geq \min_{1\leq i \leq 5}\frac{\hat Q_i}{\sup_{t,\mu\in \mathcal{P}} \{ ||\hat P_i(t,\mu)||\}} ,
  \nonumber
\end{equation} 
where $\hat Q_i$ and $\hat P_i(\cdot,\cdot) $ represent the $i$-th block diagonal element.
According to Proposition~\ref{pr-cr}, pick $\alpha_{\rm cr} = \bar a_3/2 + \phi_S$ and $P_F(t) = I$, with $||S(\cdot)||\leq \phi_S$.
Fix $K, \bar a_3$ such that $ 1/(4||P_K||), \bar a_3/2 \geq a_3/4a_2$. 
So if \eqref{eq-kg} is strengthened as
\begin{equation}
  \begin{cases}
    k& >  \phi_0(\gamma_1 + \frac{a_3}{2a_2}) \\
    g &>  \phi_0( k + \gamma_2 + \frac{a_3}{4a_2}) ,
  \end{cases} \nonumber
\end{equation}
then it is ensured $\gamma \geq a_3/4a_2$, meaning that the decay rate of the closed-loop system is restricted only by that of zero-dynamics $A_\eta(\cdot,\cdot)$. 
Hence, $ \alpha_{\rm cl}  = a_3/8a_2$.

Then considering $\phi_{\rm cl}(k_b,k_\eta)$,  
denote $\phi_{p1} = \sup_{t,\mu\in\mathcal{P}} \\\{||\tilde P^{-1}(t,\mu)||\}$ and $\phi_{p2}=\sup_{t,\mu\in\mathcal{P}}\{||\tilde P(t,\mu)||\} $,
where  $\tilde P(t,\mu) = T^T(t,\mu) \hat P(t,\mu) T(t,\mu)$.
Calculations yield 
$\phi_{p1} \leq  \phi_0  \phi_G^2  $. 
According to the Rayleigh quotient, 
$\phi_{p2} =\max_{x\neq 0,t,\mu\in\mathcal{P}} (x^T \tilde P(t,\mu) x/x^T x)\leq \phi_0 \phi_G^4  g^{2(r-1)}  $.
Since $ ||\mathbf{x}(t)||^2/\phi_{p1} \leq V(t)\leq V(t_0) \leq \phi_{p2} ||\mathbf{x}(t_0)||^2  $,
we have $||\mathbf{x}(t)|| \leq \sqrt{\phi_{p1}  \phi_{p2}}||\mathbf{x}(t_0)||$, i.e.,
$ \phi_{\rm cl}(k_b,k_\eta) = \phi_0 \phi_G^3 g^{r-1} $.

Because $||\Xi(\cdot)|| = ||S(\cdot)||$, choose $\phi_F = \phi_S+\alpha_{\rm cr}$.
%similar to the proof of Theorem~\ref{thm-infim}, 
According to Proposition~\ref{pr-bd} and recalling~\eqref{eq-sim-app}, $||\Gamma(\cdot)||,||\dot \Gamma(\cdot) ||  <\phi_G$, where
% $||H|| \le \phi_0 (\bar N_b+1)(\bar N_\eta+1)$.
$  \phi_G = \phi_0 (\bar N_b+1)(\bar N_\eta+1)\le \phi_0 (k_b+1)^N (k_\eta+1)^N$.
Then $\phi_{\rm d}=\phi_0 \phi_G $, $\phi_H =  \phi_0 \phi_G e^{(\bar a_3+\phi_S)\phi_
\delta}/\phi_{\rm uco} $. 
%Note $\phi_G<\phi_{\rm d}$, and
Pick $ k = \phi_0(2 \gamma_1 + a_3/a_2), g = \phi_0(2 k  +  \gamma_2 +a_3/2a_2)$, and  $\bar a_3= a_3/a_2$.
%which is $k= \phi_0 \phi_G^2, g = \phi_0 \phi_G^2 $.
Substitute all of them with~\eqref{eq-err-app} into~\eqref{eq-et}, yielding~\eqref{eq-app-e}.

\section*{Acknowledgment}       
This work was supported in part by the National Natural Science Foundation of China under Grants 52275564 and U24A20109.

%\section*{References}

\bibliographystyle{plain}        
\bibliography{ref}

@article{de1998zeros,
  title={Zeros of continuous-time linear periodic systems},
  author={De Nicolao, Giuseppe and Ferrari-Trecate, Giancarlo and Pinzoni, Stefano},
  journal={Automatica},
  volume={34},
  number={12},
  pages={1651--1655},
  year={1998},
  publisher={Elsevier}
}

@article{wieland2011internal,
  title={An internal model principle is necessary and sufficient for linear output synchronization},
  author={Wieland, Peter and Sepulchre, Rodolphe and Allg{\"o}wer, Frank},
  journal={Automatica},
  volume={47},
  number={5},
  pages={1068--1074},
  year={2011},
  publisher={Elsevier}
}

@book{knobloch1993topics,
  title={Topics in control theory},
  author={Knobloch, Hans W and Isidori, Alberto and Dietrich, Flockerzi},
  year={1993},
  publisher={Birkh{\"a}user Basel}
}

@phdthesis{pinzoni1989stabilization,
  title={Stabilization and control of linear time-varying systems},
  author={Pinzoni, Stefano},
  year={1989},
  school={Arizona State University}
}

@ARTICLE{marconi2013internal,
  author={Marconi, Lorenzo and Teel, Andrew R.},
  journal={IEEE Transactions on Automatic Control}, 
  title={Internal Model Principle for Linear Systems With Periodic State Jumps}, 
  year={2013},
  volume={58},
  number={11},
  pages={2788-2802},
  doi={10.1109/TAC.2013.2272137}}

@article{ichikawa2006output,
  title={Output regulation of time-varying systems},
  author={Ichikawa, Akira and Katayama, Hitoshi},
  journal={Systems \& Control Letters},
  volume={55},
  number={12},
  pages={999--1005},
  year={2006},
  publisher={Elsevier}
}

@article{zhang2006linear,
  title={The linear periodic output regulation problem},
  author={Zhang, Zhen and Serrani, Andrea},
  journal={Systems \& Control Letters},
  volume={55},
  number={7},
  pages={518--529},
  year={2006},
  publisher={Elsevier}
}

@ARTICLE{zhang2009tac,
  author={Zhang, Zhen and Serrani, Andrea},
  journal={IEEE Transactions on Automatic Control}, 
  title={Adaptive Robust Output Regulation of Uncertain Linear Periodic Systems}, 
  year={2009},
  volume={54},
  number={2},
  pages={266-278},
  doi={10.1109/TAC.2008.2010891}}

@book{isidori2003robust,
  title={Robust autonomous guidance: an internal model approach},
  author={Isidori, Alberto and Marconi, Lorenzo and Serrani, Andrea},
  year={2003},
  publisher={Springer Science \& Business Media}
}

@article{astolfi2022harmonic,
  title={Harmonic internal models for structurally robust periodic output regulation},
  author={Astolfi, Daniele and Praly, Laurent and Marconi, Lorenzo},
  journal={Systems \& Control Letters},
  volume={161},
  pages={105154},
  year={2022},
  publisher={Elsevier}
}

@inproceedings{bin2016robust,
  title={Robust internal model design by nonlinear regression via low-power high-gain observers},
  author={Bin, Michelangelo and Astolfi, Daniele and Marconi, Lorenzo},
  booktitle={2016 IEEE 55th Conference on Decision and Control (CDC)},
  pages={4740--4745},
  year={2016},
  organization={IEEE}
}

@article{serrani2002semi,
  title={Semi-global nonlinear output regulation with adaptive internal model},
  author={Serrani, Andrea and Isidori, Alberto and Marconi, Lorenzo},
  journal={IEEE Transactions on Automatic Control},
  volume={46},
  number={8},
  pages={1178--1194},
  year={2001},
  publisher={IEEE}
}

@article{paunonen2012periodic,
  title={Periodic output regulation for distributed parameter systems},
  author={Paunonen, Lassi and Pohjolainen, Seppo},
  journal={Mathematics of Control, Signals, and Systems},
  volume={24},
  number={4},
  pages={403--441},
  year={2012},
  publisher={Springer}
}

@article{weiss1968structure,
  title={On the structure theory of linear differential systems},
  author={Weiss, Leonard},
  journal={SIAM Journal on Control},
  volume={6},
  number={4},
  pages={659--680},
  year={1968},
  publisher={SIAM}
}

@article{paunonen2017robust,
  title={Robust output regulation for continuous-time periodic systems},
  author={Paunonen, Lassi},
  journal={IEEE Transactions on Automatic Control},
  volume={62},
  number={9},
  pages={4363--4375},
  year={2017},
  publisher={IEEE}
}

@article{davison1976robust,
  title={The robust control of a servomechanism problem for linear time-invariant multivariable systems},
  author={Davison, Edward},
  journal={IEEE transactions on Automatic Control},
  volume={21},
  number={1},
  pages={25--34},
  year={1976},
  publisher={IEEE}
}

@article{francis1977linear,
  title={The linear multivariable regulator problem},
  author={Francis, Bruce A},
  journal={SIAM Journal on Control and Optimization},
  volume={15},
  number={3},
  pages={486--505},
  year={1977},
  publisher={SIAM}
}

@article{francis1976internal,
  title={The internal model principle of control theory},
  author={Francis, Bruce A and Wonham, Walter Murray},
  journal={Automatica},
  volume={12},
  number={5},
  pages={457--465},
  year={1976},
  publisher={Elsevier}
}

@article{sun2009trajectory,
  title={Trajectory tracking and disturbance rejection for linear time-varying systems: Input/output representation},
  author={Sun, Zongxuan and Zhang, Zhen and Tsao, Tsu-Chin},
  journal={Systems \& Control Letters},
  volume={58},
  number={6},
  pages={452--460},
  year={2009},
  publisher={Elsevier}
}

@article{zhang2010novel,
  title={A novel internal model-based tracking control for a class of linear time-varying systems},
  author={Zhang, Zhen and Sun, Zongxuan},
  journal={Journal of Dynamic Systems, Measurement and Control, Transactions of the ASME},
  volume={132},
  number={1},
  pages={1--10},
  year={2010},
  publisher={American Society of Mechanical Engineers (ASME)}
}

@article{zhang2014discrete,
  title={A discrete time-varying internal model-based approach for high precision tracking of a multi-axis servo gantry},
  author={Zhang, Zhen and Yan, Peng and Jiang, Huan and Ye, Peiqing},
  journal={ISA transactions},
  volume={53},
  number={5},
  pages={1695--1703},
  year={2014},
  publisher={Elsevier}
}

@article{niu2024adaptive,
  title={Adaptive Observer-Based Output Regulation With Non-Smooth Non-Periodic Exogenous Signals},
  author={Niu, Zirui and Chen, Kaiwen and Scarciotti, Giordano},
  journal={IEEE Control Systems Letters},
  volume={8},
  pages={1535--1540},
  year={2024},
  publisher={IEEE}
}

@article{song2015low,
  title={Low-order stabilizer design for discrete linear time-varying internal model-based system},
  author={Song, Xingyong and Gillella, Pradeep Kumar and Sun, Zongxuan},
  journal={IEEE/ASME Transactions on Mechatronics},
  volume={20},
  number={6},
  pages={2666--2677},
  year={2015},
  publisher={IEEE}
}

@article{cao2025enhanced,
  title={Enhanced Contour Tracking: A Time-Varying Internal Model Principle-Based Approach},
  author={Cao, Yue and Zhang, Zhen},
  journal={IEEE/ASME Transactions on Mechatronics},
  year={2025},
  publisher={IEEE}
}

@article{li2025global,
  title={Global {$H_{\infty}$} Output Regulation of Nonlinear Systems Via Output-Feedback Reinforcement Learning},
  author={Li, Shaobao and Wang, Yuxiang and Zhang, Yuguang and Luo, Xiaoyuan and Wang, Juan and Guan, Xinping},
  journal={IEEE Transactions on Automatic Control},
  year={2025},
  publisher={IEEE}
}

@article{lin2025direct,
  title={Direct Adaptive Cooperative Output Regulation of Unknown Multi-Agent Systems via Distributed Internal Model},
  author={Lin, Liquan and Huang, Jie},
  journal={IEEE Transactions on Automatic Control},
  year={2025},
  publisher={IEEE}
}

@article{jiang2025output,
  title={Output Feedback-Based Adaptive Optimal Output Regulation for Continuous-Time Strict-Feedback Nonlinear Systems},
  author={Jiang, Yi and Chai, Tianyou and Chen, Guanrong},
  journal={IEEE Transactions on Automatic Control},
  year={2025},
  publisher={IEEE}
}

@article{liu2025adaptive,
  title={Adaptive-Dynamic-Programming-Regulated Extremum Seeking for Distributed Feedback Optimization},
  author={Liu, Tong and Krstić, Miroslav and Jiang, Zhong-Ping},
  journal={IEEE Transactions on Automatic Control},
  year={2025},
  publisher={IEEE}
}

@article{gul2026multi,
  title={Multi-partite output regulation of multi-agent systems},
  author={G{\"u}l, K{\"u}r{\c{s}}ad Metehan and Sars{\i}lmaz, Selahattin Burak},
  journal={IEEE Transactions on Automatic Control},
  year={2026},
  publisher={IEEE}
}

@article{li2026data,
  title={Data-Driven Cooperative Output Regulation Via Output Feedback Control},
  author={Li, Yifei and Liu, Wenjie and Wang, Gang and Xie, Lihua and Sun, Jian and Chen, Jie},
  journal={IEEE Transactions on Automatic Control},
  year={2026},
  publisher={IEEE}
}

@article{guo2025practical,
  title={Practical tracking for minimum-phase 1-{D} heat equation with arbitrary reference signals},
  author={Guo, Bao-Zhu and Wang, Ji-Xiao},
  journal={IEEE Transactions on Automatic Control},
  year={2025},
  publisher={IEEE}
}

@article{zhao2025robust,
  title={Robust output regulation for multi-dimensional heat equation under boundary control},
  author={Zhao, Ren-Xi and Guo, Bao-Zhu and Paunonen, Lassi},
  journal={Automatica},
  volume={171},
  pages={111956},
  year={2025},
  publisher={Elsevier}
}

@article{bin2022robustness,
  title={About robustness of control systems embedding an internal model},
  author={Bin, Michelangelo and Astolfi, Daniele and Marconi, Lorenzo},
  journal={IEEE Transactions on Automatic Control},
  volume={68},
  number={3},
  pages={1306--1320},
  year={2023},
  publisher={IEEE}
}

@article{bin2024robust,
  title={Robust internal models with a star-shaped attractor are linear},
  author={Bin, Michelangelo and Astolfi, Daniele and Marconi, Lorenzo},
  journal={Automatica},
  volume={166},
  pages={111698},
  year={2024},
  publisher={Elsevier}
}

@article{shim2010note,
  title={A note on the differential regulator equation for non-minimum phase linear systems with time-varying exosystems},
  author={Shim, Hyungbo and Kim, Jung-Su and Kim, Hongkeun and Back, Juhoon},
  journal={Automatica},
  volume={46},
  number={3},
  pages={605--609},
  year={2010},
  publisher={Elsevier}
}

@inproceedings{shim2006output,
  title={Output regulation problem and solution for LTV minimum phase systems with time-varying exosystem},
  author={Shim, Hyungbo and Lee, Jaehwa and Kim, Jung-su and Back, Juhoon},
  booktitle={2006 SICE-ICASE International Joint Conference},
  pages={1823--1827},
  year={2006},
  organization={IEEE}
}

@article{song2014robust,
  title={Robust stabilizer design for linear time-varying internal model based output regulation and its application to an electrohydraulic system},
  author={Song, Xingyong and Wang, Yu and Sun, Zongxuan},
  journal={Automatica},
  volume={50},
  number={4},
  pages={1128--1134},
  year={2014},
  publisher={Elsevier}
}

@article{carnevale2017robust,
  title={Robust hybrid output regulation for linear systems with periodic jumps: Semiclassical internal model design},
  author={Carnevale, Daniele and Galeani, Sergio and Menini, Laura and Sassano, Mario},
  journal={IEEE Transactions on Automatic Control},
  volume={62},
  number={12},
  pages={6649--6656},
  year={2017},
  publisher={IEEE}
}

@article{isidori1990output,
  title={Output regulation of nonlinear systems},
  author={Isidori, A and Byrnes, CI},
  journal={IEEE Transactions on Automatic Control},
  volume={35},
  number={2},
  pages={131--140},
  year={1990},
  publisher={Institute of Electrical and Electronics Engineers (IEEE)}
}

@article{pinzoni1993output,
  title={Output regulation of linear time-varying systems},
  author={Pinzoni, Stefano},
  journal={IFAC Proceedings Volumes},
  volume={26},
  number={2},
  pages={311--313},
  year={1993},
  publisher={Elsevier}
}

@article{paunonen2010internal,
  title={Internal model theory for distributed parameter systems},
  author={Paunonen, Lassi and Pohjolainen, Seppo},
  journal={SIAM Journal on Control and Optimization},
  volume={48},
  number={7},
  pages={4753--4775},
  year={2010},
  publisher={SIAM}
}

@article{cox2016isolating,
  title={Isolating invisible dynamics in the design of robust hybrid internal models},
  author={Cox, Nicholas and Marconi, Lorenzo and Teel, Andrew R},
  journal={Automatica},
  volume={68},
  pages={56--68},
  year={2016},
  publisher={Elsevier}
}

@article{berger2015zero,
  title={Zero dynamics and stabilization for analytic linear systems},
  author={Berger, Thomas and Ilchmann, Achim and Wirth, Fabian},
  journal={Acta Applicandae Mathematicae},
  volume={138},
  number={1},
  pages={17--57},
  year={2015},
  publisher={Springer}
}

\end{document}